\documentclass[
5p, 
twocolumn, 
]{elsarticle}
\usepackage{amsmath}
\usepackage{footnote}
\usepackage{threeparttable, tablefootnote}
\usepackage{tabularx}
\usepackage{bm}%
\usepackage{graphicx}
\usepackage{graphics}
\usepackage{xcolor}
\usepackage{booktabs}
\usepackage{mathrsfs}
\usepackage{caption}
\usepackage{enumitem}
\usepackage{subcaption}
\usepackage{collectbox}
\usepackage{fancyvrb}
\usepackage{multirow} 
\usepackage{hyperref}
\hypersetup{
    colorlinks,
    citecolor=black,
    filecolor=black,
   linkcolor=black,
    urlcolor=black
}
\usepackage{cuted}
\usepackage[left]{lineno} 
\setlist{nosep} 
\AtBeginDocument{
\heavyrulewidth=.08em
\lightrulewidth=.05em
\cmidrulewidth=.03em
\belowrulesep=.65ex
\belowbottomsep=0pt
\aboverulesep=.4ex
\abovetopsep=0pt
\cmidrulesep=\doublerulesep
\cmidrulekern=.5em
\defaultaddspace=.5em
}
\newcommand{\radis}{\ensuremath{ \mathrm{^{225}Ra}}}
\newcommand*\circled[1]{
\raisebox{.5pt}{\textcircled{\raisebox{-.9pt} {#1}}}
            }
\newcommand{\drplotwidth}{0.92\textwidth}

\let\today\relax
\makeatletter
\def\ps@pprintTitle{%
\let\@oddhead\@empty
\let\@evenhead\@empty
\def\@oddfoot{\textit{\footnotesize {Manuscript accepted on August 10, 2021}\hfill} {\footnotesize\url{https://doi.org/10.1016/j.nima.2021.165738}}}%
\let\@evenfoot\@oddfoot}
\makeatother



\makeatletter
\newcommand*\bigcdot{\mathpalette\bigcdot@{.5}}
\newcommand*\bigcdot@[2]{\mathbin{\vcenter{\hbox{\scalebox{#2}{$\m@th#1\bullet$}}}}}
\makeatother

\expandafter\ifx\csname package@font\endcsname\relax\else
 \expandafter\expandafter
 \expandafter\usepackage
 \expandafter\expandafter
 \expandafter{\csname package@font\endcsname}%
\fi

\begin{document}
\begin{frontmatter}
\title{Surface Processing and Discharge-Conditioning of High Voltage Electrodes for the Ra EDM Experiment}
\author[1]{Roy~A.~Ready}  
\ead{roy.a.ready@gmail.com}
\author[1]{Gordon~Arrowsmith-Kron}
\author[2]{Kevin~G.~Bailey}
\author[1]{Dominic~Battaglia}
\author[2]{Michael~Bishof}
\author[1]{Daniel~Coulter}
\author[2]{Matthew~R.~Dietrich}

\author[1]{Ruoyu~Fang}
\author[1]{Brian~Hanley}
\author[1]{Jake~Huneau}
\author[1]{Sean~Kennedy}
\author[1]{Peyton~Lalain}
\author[1]{Benjamin~Loseth}
\author[1]{Kellen~McGee}

\author[2]{Peter~Mueller}
\author[2]{Thomas~P.~O'Connor}
\author[1]{Jordan~O'Kronley}
\author[1]{Adam~Powers}
\author[1]{Tenzin~Rabga}
\author[1]{Andrew~Sanchez}
\author[1]{Eli~Schalk}
\author[1]{Dale~Waldo}
\author[1]{Jacob~Wescott}
\author[1]{Jaideep~T.~Singh}

\address[1]{National Superconducting Cyclotron Laboratory and Department of Physics and Astronomy, Michigan State University, East Lansing, Michigan 48824, USA}
\address[2]{Physics Division, Argonne National Laboratory, Argonne, Illinois 60439, USA}

\date{\today}
\begin{abstract}
\noindent
The Ra EDM experiment uses a pair of  high voltage electrodes to search for the atomic electric dipole moment of \radis. 
We use identical, plane-parallel electrodes with a primary high gradient surface of 200~mm$^2$\ to generate reversible DC electric fields.
Our statistical sensitivity is linearly proportional to the electric field strength in the electrode gap.
We adapted surface decontamination and processing techniques from accelerator physics literature to chemical polish and clean a suite of newly fabricated large-grain niobium and grade-2 titanium electrodes.
Three pairs of niobium electrodes and one pair of titanium electrodes were discharge-conditioned with a custom high voltage test station at electric field strengths as high as +52.5 kV/mm and $- 51.5$\ kV/mm over electrode gap sizes ranging from 0.4~mm to 2.5~mm.
One pair of large-grain niobium electrodes was discharge-conditioned and validated to operate at $\pm 20 \ \mathrm{kV/mm}$ with steady-state leakage current $\leq 25$\ pA ($1\sigma$) and a polarity-averaged $98 \pm 19$\ discharges per hour. 
These electrodes were installed in the Ra~EDM experimental apparatus, replacing a copper electrode pair, and were revalidated to $\pm 20 \ \mathrm{kV/mm}$. 
The niobium electrodes perform at an electric field strength 3.1 times larger than the legacy copper electrodes and are ultimately limited by the maximum output of our 30 kV bipolar power supply.
\end{abstract}

\begin{keyword}
high voltage electrode conditioning\sep leakage current  \sep large-grain niobium  \sep radium-225 \sep atomic electric dipole moment \sep magnetic Johnson noise
\end{keyword}

\end{frontmatter}


\section{Ra EDM Motivation and Requirements}

\subsection{Motivation}

Violation of combined charge conjugation ($C$) and parity ($P$) symmetry, or $CP$, is a necessary ingredient of the observed dominance of matter over antimatter, or baryon asymmetry of the universe (BAU).
$CP$ violation is encoded in the Standard Model (SM) by a complex phase term in the Cabibbo-Kobayashi-Maskawa (CKM) quark mixing matrix \cite{Ch19}. 
The SM critically underestimates the BAU, suggesting that new sources of $CP$\ violation have yet to be discovered~\cite{Hu95}. \par

Permanent electric dipole moments (EDMs) violate time-reversal ($T$) and $P$\ symmetry.
Assuming $CPT$\ conservation, EDMs also violate $CP$.
Neutron, electron, molecular, and atomic EDM experiments have been carried out over the last seven decades in an effort to measure a nonzero EDM magnitude.
While nonzero EDMs remain out of reach for now, the precision of EDM measurements continues to improve.
Observing a nonzero EDM near sensitivities of today's leading experiments would provide a clean signature of Beyond the Standard Model physics~\cite{Ch19}.
\par

The atomic EDM of \radis\ (spin $I = 1/2$) is enhanced by the octupole deformation (``pear shape'') of its nucleus. 
Radium-225 has a 55~keV parity doublet ground state structure, compared to approximately 1~MeV in spherically symmetric nuclei~\cite{Hax83}.
This enhances  the observable component of the EDM, characterized by the nuclear Schiff moment.
The Schiff moment of \radis\ is predicted to be up to three orders of magnitude larger than that of diamagnetic atoms with spherically symmetric nuclei~\cite{Aue96,Dob05,Ban10,Dzu02}.
 \par

The Ra EDM experiment (Argonne National Lab, Michigan State University) measures the spin precession frequency of \radis\ in a controlled, uniform magnetic and electric field between two high voltage electrodes in an optical dipole trap (ODT).
EDM measurements are performed at Argonne National Lab (ANL), while offline upgrades such as the high voltage development discussed in this report are carried out at Michigan State University (MSU).
In the proof of principle measurement, the EDM  $2\sigma$ upper limit was measured to \mbox{$5.0 \times 10^{-22}~e~\mathrm{cm}$~\cite{Pa15}.}
This was reduced to \mbox{$1.4 \times 10^{-23}~e~\mathrm{cm}$} in the subsequent run~\cite{Bi16}.
Hereafter we will refer to these as the `first generation' measurements.  
\par

The shot noise-limited EDM standard error \mbox{$\large{\sigma}_{\mathrm{EDM}}\ (e~\mathrm{cm})$} is given by:
\begin{equation} \label{statistical-sensitivity}
\large{\sigma}_{\mathrm{EDM}} = \dfrac{\hbar}{2E\sqrt{\epsilon N T \tau \ }} \rlap{\,,}
\end{equation}
\begin{description}
\item where\\
$ E$\ $(\mathrm{V/cm})$\ is the external electric field,\\
$ \hbar$\ $(\mathrm{eV \ s})$\ is the reduced Planck constant,\\
$ \epsilon$\ $(\mathrm{unitless})$\  is the atom detection efficiency,\\
$N$\ $ (\mathrm{unitless} )$\ is the number of atoms per sample,\\
$T$\ $ (\mathrm{s}) $\ is the total measurement time, and \\
$\tau $\ $ (\mathrm{s} )$\ is the measurement time per cycle.\\
\end{description}

\noindent As seen in Equation~\ref{statistical-sensitivity}, the statistical sensitivity of the EDM measurement scales linearly with the electric field strength.
The Ra~EDM experiment will be significantly improved with targeted upgrades to the experimental apparatus over the next several `second generation' measurements.
In particular, we will use a new atom detection method to increase $\epsilon$\ and new electrodes to increase $E$.
We will surpass the \mbox{$10^{-25} \ e$ cm}  sensitivity level during this phase and the \radis \ EDM limit will constrain hadronic $CP$-violating parameters alongside other EDM experiments~\cite{Ch15}.

\begin{figure}
\centering
\includegraphics[width=0.40\textwidth]{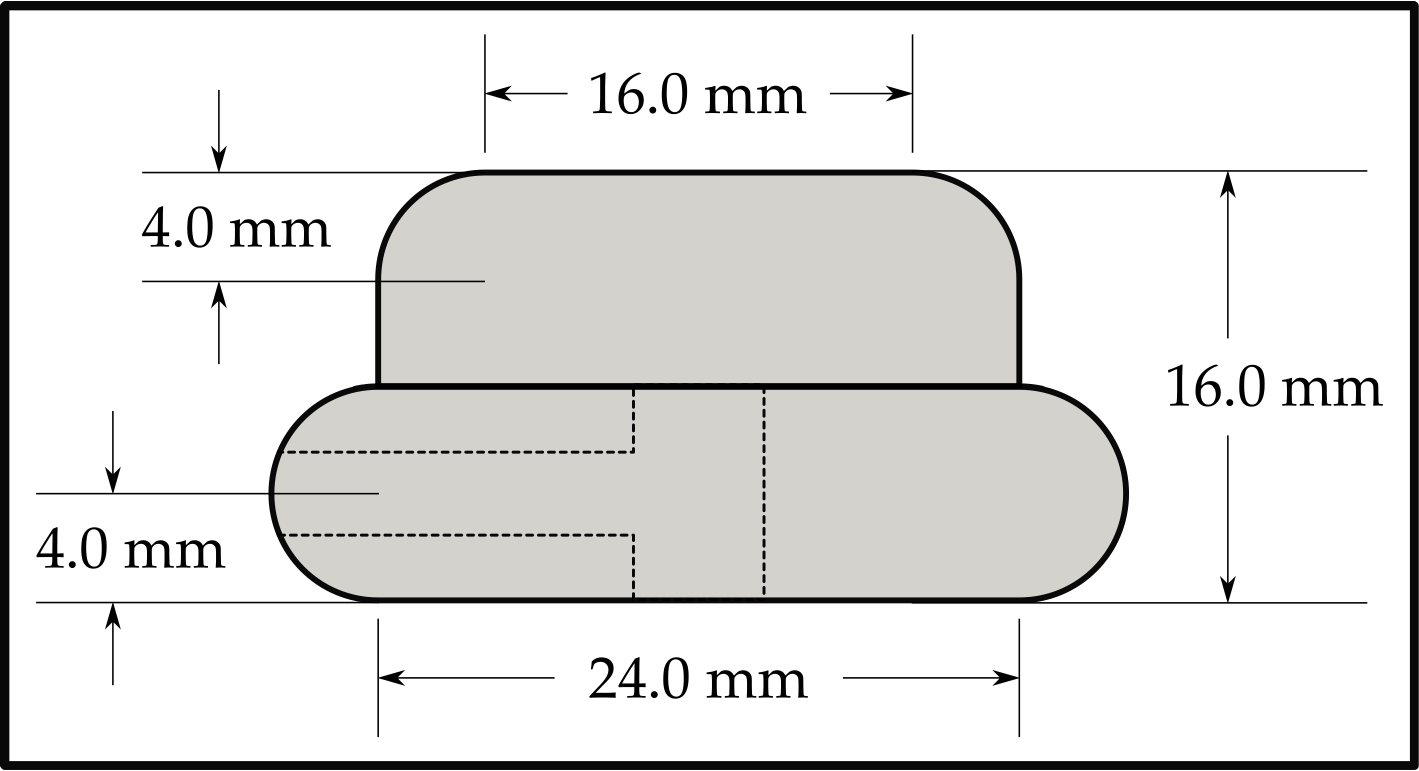}
\caption{\label{electrode-geometry}
Cross-sectional electrode schematic. 
Surfaces have a  flatness tolerance of 25.4~$\mu$m and a parallelism of 50.8~$\mu$m. 
The top surface is polished to an average roughness of 0.127~$\mu$m.
The base is mounted by a 10-32 tapped hole.
}
\end{figure}

\subsection{\label{experiment-requirements}High voltage experimental requirements}

The EDM couples to an external electric field analogously to the coupling of the atomic magnetic dipole moment to an external magnetic field.
The Hamiltonian of an atom in the presence of a perfectly uniform electric and magnetic field is given by the following:

\begin{equation}
\label{hamiltonian}
\mathcal{H} = -\mu\left( \frac{ \vec{S} \cdot \vec{B} }{S} \right) -d\left( \frac{\vec{S} \cdot \vec{E}}{S} \right) \rlap{\,,}
\end{equation}
\begin{description}
 \item where \\
$\mu = -2.3 \times 10^{-8}$\ eV/T is the atomic magnetic dipole moment of $^{225}$Ra~\cite{Arn87}, \\
$\vec{S}$ is the total atomic angular momentum, \\
$\vec{B}$\ (T) is the applied magnetic field, \\
$d \ (e$ cm) is the atomic EDM, and \\
$\vec{E}$\ (V/cm) is the applied electric field.
\end{description}

\noindent The \radis \ atoms will precess with frequency $\omega_{+}$\ $(\omega_{-})$\ when $\vec{E}$ is parallel (antiparallel) to $\vec{B}$:
\begin{equation}
\label{frequency}
\omega_{\pm} = \frac{2}{\hbar}(\mu B \pm dE) \ \mathrm{rad/s}
\end{equation}
\noindent In the most recent Ra~EDM experiment we applied a \mbox{$2.6~\mu$T} magnetic field and measured a spin precession frequency of \mbox{$181.1 \pm 1.6$\ rad/s \cite{Bi16}.}

We use a pair of identical plane-parallel electrodes to produce a stable, uniform, and symmetric electric field.
The spin precession of the atoms is measured in three configurations: with the electric field parallel to the magnetic field, with the electric field antiparallel to the magnetic  field, and with no applied electric field.
The ``field-off'' setting is used to control for a systematic effect generated by an imperfect reversal of the electric field.
We measure the accumulated spin precession phase for each field configuration.
The extracted EDM is related to the accumulated phase difference between the parallel and antiparallel configurations by Equation~\ref{edm}:
\begin{equation}
\label{edm}
d = \frac{\hbar \Delta \phi}{4 E \tau} \rlap{\,,}
\end{equation}

\noindent where $\Delta \phi$ (rad) is the difference in accumulated phase between the two ``field-on'' configurations. 
With a perfectly uniform and static magnetic field under all configurations, the phase difference between the parallel and antiparallel field configurations is purely due to the EDM interaction with the electric field.
A higher electric field strength will generate a larger accumulated phase and improve EDM sensitivity. \par

\begin{table}
\centering
\begin{threeparttable}
	\caption{\label{raedm} Ra EDM systematic requirements at the \mbox{$10^{-26} \ e$ cm} sensitivity level.
	Detailed descriptions of evaluations of $\theta_{E},\ |\Delta E| / E,\ \mathrm{and} \ \bar{I}$\ can be found in our previous work~\cite{Bi16}.
	$\Delta B$\ is determined by Equation~\ref{false-edm}.
	We describe our calculation of the Johnson noise limit in \ref{mjn-calc}.
	}
\begin{tabular}{@{}lcl@{}}
\toprule
description & \multicolumn{2}{c}{systematic limit}
\\
\midrule
\vspace{0.5em}
$\vec{E} , \ \vec{B} $\ alignment &   $\theta_{E}$  &  $\leq$\ 2 mrad \\\vspace{0.5em}
polarity imbalance  & $\dfrac{|\Delta E|}{E}$ &  $\leq$\ 0.7\% \\\vspace{0.5em}
electrode magnetic impurity  & $\Delta B$ &  $\leq \ 100 \ \mathrm{fT}$\tnote{a} \\\vspace{0.5em}
steady-state leakage current  & $ \bar{I}$ & $\leq \ 100 \ \mathrm{pA}$\tnote{a}\\\vspace{0.5em}
magnetic Johnson noise  & $ \sqrt{\dfrac{dB^2_n}{d\nu}}$ & $\leq 15 \ \dfrac{\ \mathrm{pT}\tnote{a}}{\sqrt{\mathrm{Hz}}}$  
\\
\bottomrule
\end{tabular}
\begin{tablenotes}\footnotesize
\item [a] per measurement cycle
\end{tablenotes}
\end{threeparttable}
\end{table}

In one measurement cycle, one electrode is charged to \mbox{$\leq +30$~kV} (positive polarity) while the other is grounded. 
The atom trap lifetime is currently about twenty seconds.
We expect to increase the trap lifetime to one hundred seconds~\cite{Romalis1999} as improvements are made to the ODT.
The charged electrode is then ramped to zero voltage and remains grounded for a period of  60~s while a new sample of atoms is prepared.
The next cycle begins and the electrode is charged to the same voltage magnitude at negative polarity.
We repeat this process until the atomic oven is depleted after approximately two weeks. \par

Now we will discuss EDM measurement systematics related to the high voltage system.
Our requirements for each systematic are given in Table~\ref{raedm}.\par

\begin{figure}
\centering
\includegraphics[width=0.45\textwidth]{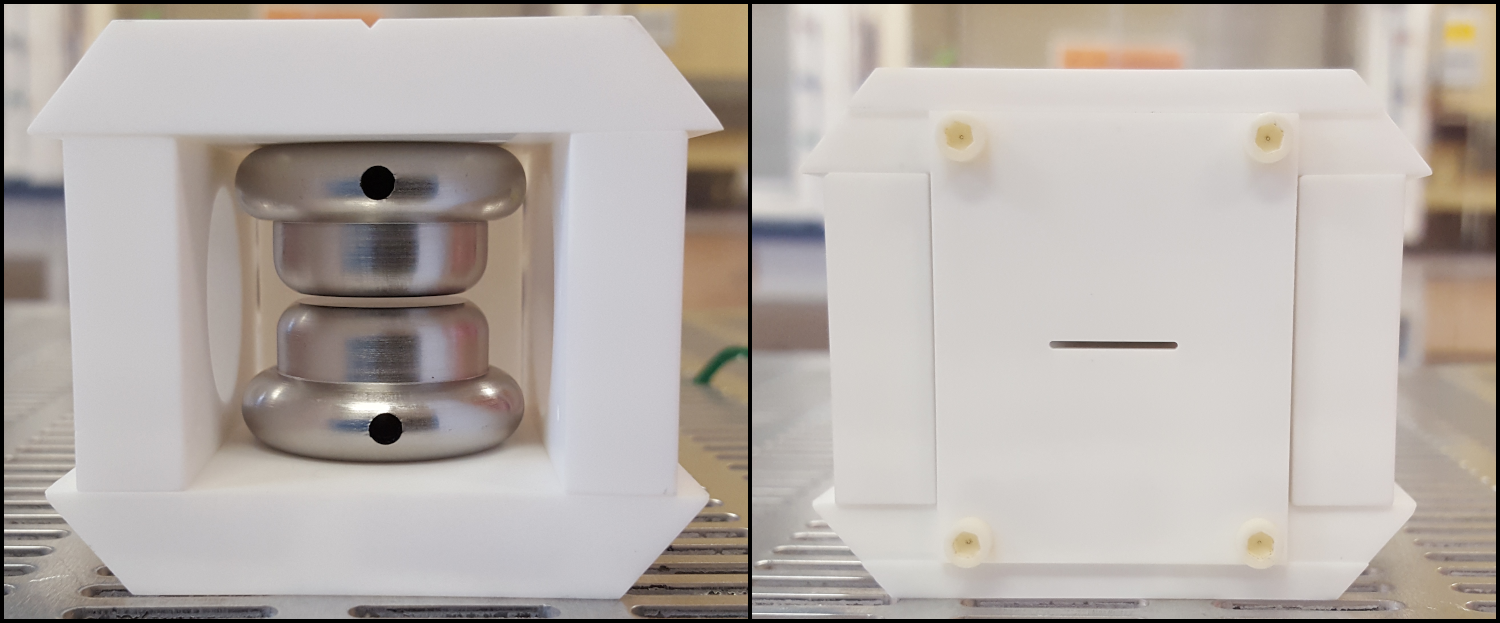}
\caption{\label{niobium-in-holder}Left: assembly of the niobium pair Nb$_{56}$ at 1 mm gap in Macor holder. 
Right: a slit centered on the gap shields the electrode surfaces from heating by the atom-trapping and excitation lasers.  }
\end{figure}

The electric field between the electrodes must be symmetric, uniform, and reversible to minimize systematic effects. 
The alignment between $\vec{E}$\ and $\vec{B}$\ is fixed after mounting the electrodes to the Macor holder, as shown in Figure~\ref{niobium-in-holder}. 
In the experimental apparatus, the holder and electrodes rest within a borosilicate glass tube.
We will use vector fluxgates with a system of autocollimators to optically determine the field uniformity and alignment for the second generation EDM measurements~\cite{Di17}. 
The field reversibility is measured with a calibrated high voltage divider (Ross Engineering V30-8.3-A). \par

Magnetic field fluctuations caused by current in the electrodes, or magnetic Johnson noise, limits the choice of electrode materials and geometries that are suitable for an EDM measurement.
The magnetic field scales as $\rho^{-1/2}$, where \mbox{$\rho$\ ($\Omega \ $m)} is the resistivity.
For two niobium electrodes separated by 1~mm with the geometry shown in Figure~\ref{electrode-geometry}, we estimate the magnetic Johnson noise per sample to be 2.48 pT/$\sqrt{\mathrm{Hz}}$. 
For an EDM measurement  lasting $T=15$\ days with an atom spin precession time of \mbox{$\tau = 100$\ s} and an electric field of 30~kV/mm (see Equation~\ref{statistical-sensitivity}), magnetic Johnson noise will only become significant at the \mbox{$10^{-26}~e$~cm} level.
A detailed description of magnetic Johnson noise calculations is given in \ref{mjn-calc}.
\par

We consider an additional systematic in which the magnetization of a fraction of the impurities in the electrodes depends on the polarity of the charging current.
A sufficiently high concentration of paramagnetic impurities near an electrode primary surface could perturb the magnetic field in the radium cloud region. 
This would generate an atomic precession frequency mimicking an EDM signal, which can be expressed as a ``false'' EDM $d_{\Delta B}$:

\begin{equation}
\label{false-edm}
d_{\Delta B} = \frac{\mu \Delta B}{E} \rlap{\,,}
\end{equation}
\noindent where $\Delta B$\ is the local magnetic field change from magnetic impurities in the electrodes as the electric field is reversed. 
\par

For a local magnetic field change \mbox{$\Delta B \approx 100\ \mathrm{fT}$} per \mbox{30 kV/mm} field reversal, this systematic will only become significant at the \mbox{$10^{-26}~e$~cm} level.
Measuring a magnetic field strength of this magnitude will require more sensitive techniques than the low-noise fluxgate magnetometers (Bartington Mag-03MSL70) we currently use. 
\par

To minimize systematic effects due to magnetic impurities, we use high-grade electrode materials and surface processing techniques that remove contaminants.
Tables~\ref{bulk}~and~\ref{electrodes} list the material properties and processing techniques that we use.
We will discuss electrode material selection and surface processing in detail in Section~\ref{section-electrode-properties-prep}.

\subsection{Electric field and laser interactions} \label{apparatus}

A radium sample is trapped in the electrode gap by an ODT for each EDM measurement cycle.
We induce coherent atomic spin precession in the controlled magnetic and electric field with a polarizing laser pulse.
The spin precession frequency is measured by firing a subsequent `detection' laser pulse and imaging the atom cloud photon absorption fraction after a variable delay time $\delta$~(ms).
\par

\begin{figure}
	\centering
	\includegraphics[width = 0.36\textwidth]{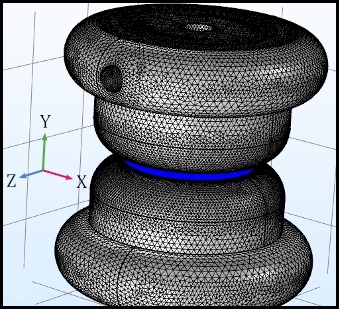}
	\caption{\label{comsol-model}A  COMSOL meshed model of the electrodes with the region of interest shaded blue.
	The origin is 0.5 mm below the top electrode surface, centered on the top electrode.
	When the bottom electrode is aligned with the top electrode, the origin is exactly in the center of the electrode gap.}
\end{figure}

We considered two methods for the sequencing of the polarization pulse, detection pulse, and the electric field ramping.
In the first method, the polarizing laser pulse is fired after the electric field ramps on and the detection pulse is fired before the field ramps off.
This shifts the $^{225}$Ra ground state due to an interaction between the ODT polarization and the DC electric field.
In the second method, the polarizing laser pulse is fired before ramping the field on and after ramping the field off.
This pulse sequence avoids potential mixing of the excited state hyperfine levels and suppresses atomic polarization from ODT and electric field interactions.
\par

The Ra EDM experiment uses the second method to measure the spin precession frequency.
We also consider spin precession perturbations caused by transient magnetic fields that are generated during electrode charging.
This effect is suppressed if the ramping on and ramping off pulse shapes are symmetric. 
Even with zero charging field cancellation, this systematic will only become significant at the \mbox{$10^{-27} \ e$\ cm} level~\cite{Bi16}.

\subsection{Electrode geometry}

Two identical electrodes make up the Ra~EDM electrode pair. 
The primary surface, seen as the top surface in Figure~\ref{electrode-geometry}, is flat and 16~mm in diameter.
The rounded edges have 4~mm circular radial curvatures.
We use plane-parallel electrodes (see Figure~\ref{niobium-in-holder}) so that the reversible field is uniform and symmetric as the electrodes alternate roles as cathode and anode every EDM measurement cycle.
\par

The Ra~EDM experiment requires an applied electric field that is symmetric, uniform, and reversible in the center of the electrode gap where the spin precession frequency of the \mbox{50 $\mu$m} diameter radium cloud is measured. 
Our electrode geometry reliably meets these requirements at field strengths of \mbox{12--30 kV/mm}.
In Section~\ref{comsol-field-simulation} we will use finite element modeling to show that the electric field generated by our electrodes matches that of the ideal infinite-plane capacitor in the atom cloud region.
 \par
\begin{figure}
	\centering
	\includegraphics[width=0.4825\textwidth]{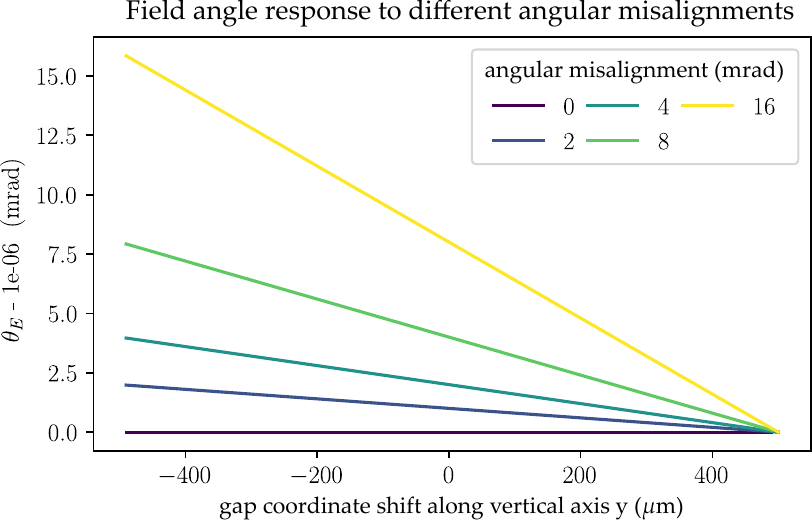}
	\caption{\label{xy-z-000-subplot-0000-offset} $\theta_E$\ as a function of the vertical distance $y$\ when the electrodes are axially aligned for a range of tilts.
	At $y=0$, $\theta_E$\ is evaluated 0.5 mm below and centered on the top electrode. }
\end{figure}
Systematic effects arising from asymmetric field reversal must continue to be reduced as EDM statistical sensitivity improves.
In the current measurement scheme, one electrode is permanently grounded and the other electrode is charged by a bipolar power supply.
We will design a more symmetric apparatus that allows us to alternate the charged and grounded electrodes using high voltage switches and a unipolar 50~kV power supply in the next phase of high voltage development.
In addition, we will optimize the electrode geometry to reduce field edge effects using the computational modeling described in Section~\ref{comsol-field-simulation}.

\subsection{\label{comsol-field-simulation} Field angle response to electrode misalignment}

One  systematic that creates a ``false'' EDM-like signal scales with the sine of the angle between the electric field and the controlled uniform magnetic field we use for measuring the spin precession frequency of the radium atoms.
We modeled the high voltage electrodes in the finite element analysis software \Verb|COMSOL Multiphysics| (version 5.3) to study the electrostatic behavior as the alignment is varied from perfectly parallel, axially-centered electrodes.
In the model, the electrodes are surrounded by a perfect vacuum.
The electrode gap size is fixed at 1~mm and the top electrode is charged to \mbox{$-30$\ kV} for a nominal electric field of \mbox{$E_0 = 30$~kV/mm.}
\par

Our simulations use the \Verb|Extremely Fine|  settings with \Verb|Size Expression| increased to $4 \times 10^{-4}$\ in the gap region and \Verb|Resolution| increased to 200 along the upper curved electrode surface.
We reduced the minimum mesh element size  to 20 $\mu$m, where we found that the electric field dependence on the mesh size converges to negligibly small fluctuations.
\par

The coordinate system of the electrode pair electrostatic model is shown in Figure~\ref{comsol-model}, with the origin defined as the midpoint between the two electrodes along their vertical axis of the top electrode.
We find that the vertical field strength $E_y $ changes by  less than 6~ppb per 100~$\mu$m when the electrodes are perfectly aligned.
The horizontal field magnitude $E_{\perp} = \sqrt{E_x^2+E_z^2\ }$\ changes by less than 5~ppb per 100~$\mu$m with respect to $E_0$ within 0.5~mm of the origin.
In practice, we  align our electrodes to better than 4~mrad in the high voltage test stand described in Section~\ref{High voltage test apparatus}.
\par

\begin{figure}
	\centering
	\includegraphics[width=0.4825\textwidth]{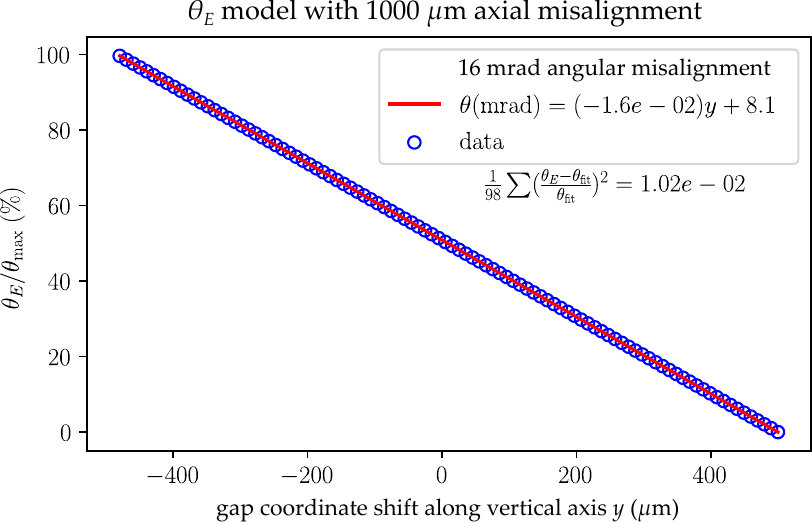}
	\caption{\label{efield-fits} A linear fit to the simulated electric field polar angle with the bottom electrode tilted 16~mrad and shifted 1~mm.
	At $y=0$, $\theta_E$\ is evaluated 0.5~mm below and centered on the top electrode.}
\end{figure}

\begin{table*}[ht!]
\centering
\begin{threeparttable}
\caption{\label{bulk} Bulk material properties of electrodes.}
\begin{tabular}{@{}llllllll@{}} 
\toprule
material & \emph{Z} \ \ & $\phi$ (eV) & strong magnetic & density  & resistivity  & hardness & outgas rate\\ 
    & 
    & 
    & 
 impurity (\%)\tnote{a}& 
(kg/$\mathrm{m^3})$ & 
($\mu \Omega \ \mathrm{cm})\tnote{b}$ 
 & 
(kgf/$\mathrm{mm^2}$) &
(Torr nL s$^{-1}$\ cm$^{-2}$)
\\ 
\midrule
niobium\tnote{c}   & 41  & 4.3 & $2.7\times 10^{-2}$ & 8570  & 15.2   &   $134.6$ & $ 30$ 
\\
copper\tnote{d}   & 29   & 4.65 & $2.5\times 10^{-7}$  & 8960 & 1.543  &  $35.0$ & $16.3$
\\
titanium\tnote{e}    & 22  & 4.33 & $5.5\times 10^{-1} $ & 4506  & 39 &   $99.0$ &   $184$
\\
stainless steel\tnote{f} & -  & 4.34 & $8.1\times 10^{+1}$  & 8000  & 69.0  &   $176$  &   $42.8$
\\
molybdenum\tnote{g}   & 42    & 4.6 & $1.4\times 10^{-2}$  & 10200 & 4.85 &   $156.0$  &   $36.7$
\\[0.5em]
\multicolumn{8}{c}{References} 
\\ 
   &     &\cite{Mi77,Wi66} &  \cite{Ha14}  & \cite{Ha14,Da98}  & \cite{Rum20,asm10}  & \cite{Sa68,Da98} & \cite{Oh03,Di98}
\\  
\bottomrule
\end{tabular}
\begin{tablenotes}
\footnotesize{
\item[a]{We define ``strong magnetic impurities'' as $\chi _{m} / (10^{-6} \ \mathrm{cm}^3 \ \mathrm{mol}^{-1}) > +1000$, where $\chi _{m}$ is the molar susceptibility. 
\mbox{$\chi_{m}(\mathrm{Nb}) = +208$.}}
\item[b]{Resistivity measured at 273 K.}
\item[c]{Hardness for at 473 K. Outgas rate estimated from the correlation between Cu, SS, and Nb desorption. }
\item[d]{Hardness of single crystal  (III) at 293 K. Outgas rate for unbaked OF high-conductivity after ten hours.}
\item[e]{Hardness of iodide-annealed, 99.99\% purity at 293 K. Outgas rate for unbaked OF high-conductivity after ten hours.}
\item[f]{Hardness of designation type 304. Outgas rate for unbaked, electropolished NS22S after ten hours.}
\item[g]{Hardness measured at 293 K.}
}
\end{tablenotes}
\end{threeparttable}
\end{table*}

 \begin{sloppypar}
We investigated the effect of misalignments between the electrodes on the electric field angle, defined as \mbox{$\theta_{E} = \arctan{\left(E_y/E_{\perp}\right)}$}.
There are two types of misalignments we consider.
Angular misalignments, or tilts, are introduced by rotating the bottom electrode about the $z$~axis in the range \mbox{0--16 mrad}.
Axial misalignments, or shifts, translates the bottom electrode along the $x$~axis and offsets the electrode centers.
Shifts of up to 1~mm displacements are considered in this work.
When the tilt and shift are zero, the electrodes are perfectly aligned and \mbox{$\theta_E$ = 0}\ near the center of the gap, corresponding to a uniform vertical field.
\end{sloppypar}

The electric field angle  scales linearly with the angular misalignments, as shown in Figure~\ref{xy-z-000-subplot-0000-offset}.
We modeled the change in $\theta_E$\ as a linear function of the position in both the $xy$\ plane (Figure~\ref{efield-fits})  and the $xz$\ plane. 
The linear model reproduces the change in the electric field angle to an accuracy of better than \mbox{$ 1\ \mu$rad} in both planes up to 1~mm from the center of the gap, even for large angular and axial misalignments. 
\par

The vertical field strength is reduced minutely even for the severe 16~mrad tilt and 1~mm shift we have modeled in Figure~\ref{efield-fits}.
We find the vertical field strength fractional change \mbox{$\Delta E_y / E_0 \approx 230 \ \mathrm{ppm}$}\ per 500~$\mu$m from the origin. 
The electrode shift effectively changes the gap size near the origin, causing a constant offset in the vertical field strength.
For the case of a 16~mrad angular misalignment and 1~mm axial misalignment, the offset in $E_y$\ is $1.6\%$.
\par

We show in Figure~\ref{efield-fits} that the field angle is described by a linear function of the vertical ($y$) coordinate.
Initially vertical ($\theta_E = 0$) at the top surface of the electrode, the field angle changes by $1\%$\ of the electrode tilt per \mbox{10 $\mu$m} along the $y$~axis.  
The field angle is 8~mrad at the midplane halfway between the electrodes and 16~mrad at the surface of the bottom electrode.
If we scan horizontally in the midplane along the $x$~axis towards the electrode edge, the polar angle changes by $0.03\%$\ per 10~$\mu$m. 
\par

In the more realistic case of a 2~mrad tilt, we find that $\theta_E$\ changes by 0.2~$\mu$rad per~100~$\mu$m in the vertical plane and 0.02~$\mu$rad per~100~$\mu$m in the midplane.
EDM systematic effects arising from field angle changes of this magnitude are far below our current statistical sensitivity.

\subsection{High voltage upgrade strategy and results\label{section:strategy-results}}

We define discharge-conditioning as the process of applying iteratively higher voltages to the electrodes to suppress steady-state leakage current and discharge rates between them. 
Leakage current refers to any current flowing between the electrodes detected by a picoammeter in series with one of the electrodes, as shown in Figure~\ref{voltage-schematic}.
We differentiate our method from the standard ``current-conditioning'' method~\cite{La95} because we characterize electrode performance by counting discrete discharges over time and we use a periodic voltage waveform.
In this paper we will interchangeably use the shorthand term ``conditioning'' when referring to discharge-conditioning. \par

In the absence of surface particulate contamination, electrode discharges are caused by charge buildup on microprotrusions on the electrode surfaces~\cite{Bo95}, which we will refer to as charge emitters.
We process and handle our electrodes in Class~100 or better environments to minimize particulate contamination.
The height of charge emitters have been measured on the order of 1~$\mu$m in buffer chemical-polished large-grain niobium electrodes prepared similarly to our electrodes~\cite{Ba12}.
If the charge emitter is near the edge of the electrode, we expect the higher gradients will increase the likelihood of a discharge. 
\par

Controlled discharges electrically polish away, or ablate charge emitters over time, allowing the electrodes to perform reliably at higher voltages~\cite{La95}. 
As shown in Section~\ref{section-testing-the-electrodes}, it may take tens to more than one hundred hours of discharge-conditioning to suppress charge emitters.
We expect the required conditioning duration may take longer if the surface is insufficiently polished or contaminated.
Bulk properties, such as the work function, resistivity, or hardness of the electrode may also play a role in the conditioning time. 
These bulk properties are listed for a selection of commonly used electrode materials in Table~\ref{bulk}. 
\par

\begin{table*}[ht!]
\centering
\begin{threeparttable}
\caption{\label{electrodes} Ra EDM electrode inventory. 
The large-grain (LG) niobium electrodes have a residual resistance ratio \mbox{(RRR) $>250$.}
\mbox{OF = oxygen} free.
\mbox{G2 = grade-2.}
Simichrome polish by hand.
Diamond paste polish (DPP) by hand.
\mbox{LPR = low} pressure rinse.
\mbox{HPR = high} pressure rinse.
\mbox{HF = hydrofluoric} chemical polish.
\mbox{EP = electropolish.}
\mbox{BCP= buffered} chemical polish.
\mbox{SiC = silicon} carbide machine polish.
\mbox{CSS = colloidal} silica suspension machine polish.
\mbox{VB = 420--450 $^{\circ}$C} vacuum outgas bake.
\mbox{WB = 150--160 $^{\circ}$C} water bake.
\mbox{USR = ultrasonic} rinse after detergent bath.
}
\begin{tabular}{@{}cllrlllll@{}}
\toprule
\multicolumn{1}{l}{batch} & 
\multicolumn{1}{l}{material} & 
\multicolumn{1}{l}{pair} & 
\multicolumn{5}{c}{surface processing recipe} 
\\ 
\midrule
1   & OF copper     & $\mathrm{Cu_{12}}$\tnote{a}  & Simichrome  & $\rightarrow$\ EP                            & $\rightarrow$ USR                                 &  $\rightarrow$\ WB                                                                               & 
\\
2   & LG niobium  & $\mathrm{Nb_{14}}$                               & SiC                  & $\rightarrow$ BCP                             & $\rightarrow$ DPP                                                & $\rightarrow$ CSS   & $\rightarrow$ USR & $\rightarrow$ VB $\cdots$
\\
     &                        &                                                                     &  $\cdots$ LPR                                                                                &   $\rightarrow$ HPR &                 &                       &                                         
\\
2   & LG niobium  & $\mathrm{Nb_{23}}$                               & SiC                   & $\rightarrow$ BCP                            & $\rightarrow$ USR &  $\rightarrow$ VB                    &  $\rightarrow$ HPR   & $\rightarrow$ resurface $\cdots$                           
\\
     &                       &                                                                      &  $\cdots$          BCP &        $\rightarrow$ HPR     &                                             &      &
\\
2   & G2 titanium & $\mathrm{Ti_{24}}$                                  & SiC                    & $\rightarrow$ HF                              & $\rightarrow$ USR & $\rightarrow$ VB  & $\rightarrow$ HPR                                                
\\
2   & G2 titanium & $\mathrm{Ti_{13}}$                                  & SiC                    & $\rightarrow$ HF                                & $\rightarrow$ EP   & $\rightarrow$ USR                                               & $\rightarrow$ VB & $\rightarrow$ HPR
\\
3   & LG niobium & $\mathrm{Nb_{56}}$\tnote{b}                               & SiC                     & $\rightarrow$ BCP                           & $\rightarrow$ USR & $\rightarrow$ HPR                       &  $\rightarrow$ WB 
\\
3   & LG niobium & $\mathrm{Nb_{78}}$                               & SiC                     & $\rightarrow$ BCP                           & $\rightarrow$ USR & $\rightarrow$ HPR                                                                                                                                    &
\\
\bottomrule
\end{tabular}
\begin{tablenotes}
\footnotesize{
\item[a]{Legacy electrodes used for first two measurements \cite{Pa15,Bi16}.}
\item[b]{Second generation electrodes described in this work and currently installed in the Ra EDM apparatus.}
}
\end{tablenotes}
\end{threeparttable}
\end{table*}

Four pairs of niobium electrodes and two pairs of titanium electrodes were surface processed as described in Table~\ref{electrodes}.
After high-pressure rinsing they are preserved in clean room environments of Class~100 (ISO 5) or better.
We conditioned pairs of electrodes in a custom, Class~100-rated high voltage test station at MSU by applying DC voltages as high as \mbox{$\pm 30$\ kV} at gap sizes in the range \mbox{0.4--2.5 mm.}
Maximum fields of \mbox{$+52.5$\ kV/mm} and \mbox{$-51.5$\ kV/mm} were tested and are discussed in Section~\ref{nb78-analysis}.
\par

One pair of large-grain niobium electrodes was validated to operate reliably at 20~kV/mm at MSU.
The electrodes were mounted in a stainless steel container and sealed in tubing backfilled with particle-filtered, dry nitrogen and were transported to ANL.
We then constructed and validated a Class~100 clean room that covered the electrode entry point to the Ra~EDM experimental apparatus.
The electrodes were removed from their packaging and installed in the apparatus  in May 2018, where they were revalidated to 20~kV/mm. \par

In Section~\ref{section-electrode-properties-prep} we will describe our past and present considerations in electrode material and surface processing.
We start by describing the preparation of the previous electrode pair used for the first generation EDM measurements in Section~\ref{early-anl}.
Material selection, surface processing, and electrode decontamination for the new electrodes tested in this work are detailed in \mbox{Sections \ref{subsection-electrode-material} and \ref{subsection-processing-techniques}.}
We will present our method of benchmarking the performance of the electrodes in Section~\ref{section-testing-the-electrodes}.
Finally, we will summarize our findings and outline a roadmap for future Ra~EDM high voltage development in Section~\ref{section-conclusions}.

\section{\label{section-electrode-properties-prep}Electrode Properties and Preparation}

\subsection{\label{early-anl}Legacy electrode preparation}

The first generation EDM measurements used a pair of electropolished oxygen-free copper electrodes~\cite{Pa15,Bi16}. 
Their geometry is identical to the new electrodes discussed in this work (Figure~\ref{electrode-geometry}).
Surface processing of these electrodes, labeled as Cu$_{12}$, is detailed in Table~\ref{electrodes}.
\par

The legacy electrodes were conditioned at ANL with a unipolar \mbox{$-30$\ kV} power supply (Glassman PS/WH-30N15-LR) in a Macor holder at a \mbox{2 mm} gap size in 2008~\cite{Graner2009}.
The electric field was reversed by turning the system off and manually switching the power supply terminations at the high voltage feedthroughs. 
Voltage was increased from 1--20~kV in 1~kV steps while monitoring the steady-state leakage current.
Conditioning was declared complete if the electrodes could hold 20~kV with a steady-state leakage current of \mbox{$< 100\ \mathrm{pA}$} for ten hours.
\par
Four pairs of electrodes total were tested in this manner, including two pairs of titanium electrodes and one pair of copper electrodes without electropolishing.
The legacy titanium electrodes all exhibited leakage current higher than 100~pA  at \mbox{20 kV}.
Flooding the test chamber with argon gas and plasma discharge-conditioning the titanium electrodes was attempted without an observable benefit. 
Both copper electrode pairs were conditioned, with the electropolished (EP) electrodes taking significantly less time.
\par

The legacy electrode pair Cu$_{12}$ was mounted in a Macor holder at a 2.3~mm gap size and installed in the \mbox{Ra EDM} experimental apparatus~\cite{Par15}. 
The pair was  retested at \mbox{20 kV / 2.3 mm = 8.7 kV/mm} but exceeded the 100~pA limit.
This was remedied by reducing the electric field by 25\%  to 6.5~kV/mm for the EDM measurement.
We suspect that the primary surface of one or both of these legacy electrodes was contaminated during installation. 
This was a motivating factor in the development of the decontamination techniques for the new electrodes discussed in subsequent sections.

\begin{figure*}
\centering
\includegraphics[width=\textwidth]{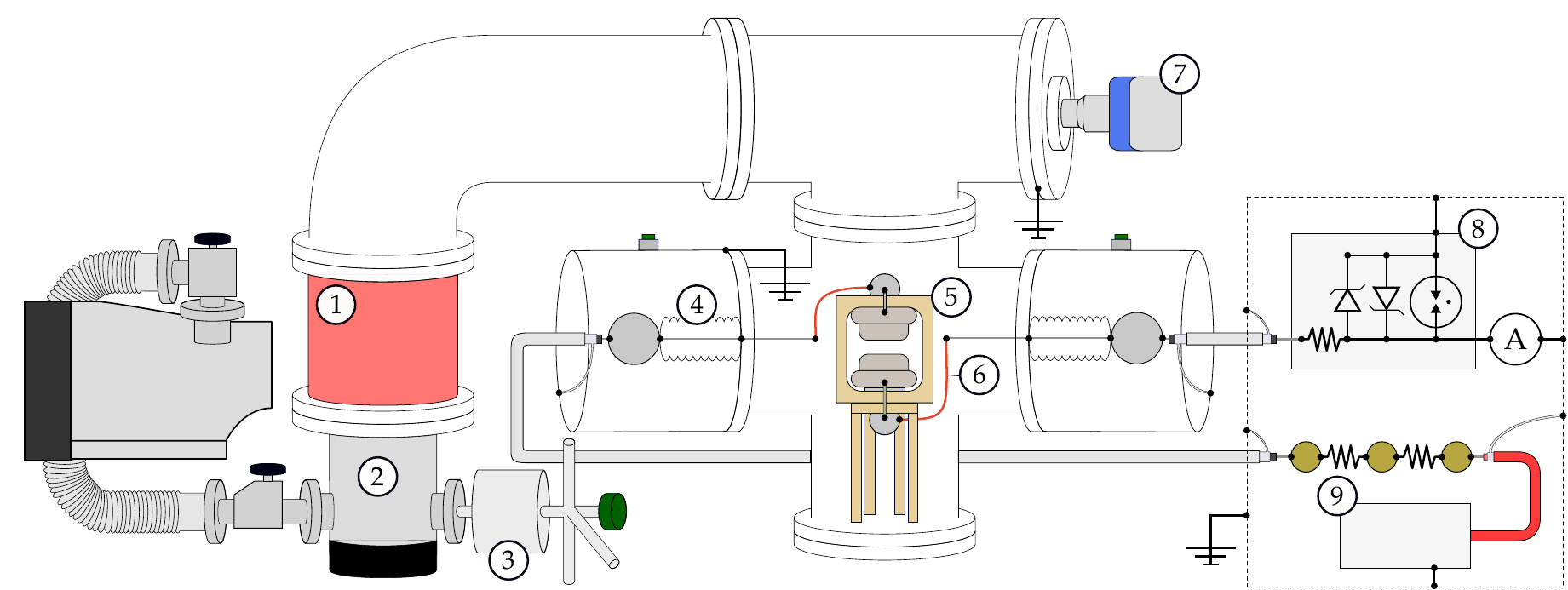}
\caption{\label{voltage-schematic}
	MSU HV test apparatus. 
		\mbox{\circled{1} 9699334} Agilent Turbo-V vibration damper
		\mbox{\circled{2} Pfeiffer} HiPace 80 turbomolecular pump with foreline Edwards~nXDS10i A736-01-983 dry scroll rough pump and two valves
		\mbox{\circled{3} Matheson} 6190 Series 0.01 micron membrane filter and purge port 
		\mbox{\circled{4} Ceramtec} 30~kV 16729-03-CF HV feedthroughs
		\mbox{\circled{5} $ 0.312 \ \mathrm{in.}^2$ electrodes} in PEEK holder \mbox{(resistivity $10^{16} \ \mathrm{M\Omega~cm}$)}
		\mbox{\circled{6} 20 AWG} Kapton-insulated, gold-plated copper wire
		\mbox{\circled{7} MKS 392502-2-YG-T}  all-range conductron/ion  gauge
		\mbox{\circled{8} Shielded} protection circuit: \mbox{Littelfuse SA5.0A} transient voltage suppressor, \mbox{EPCOS EX-75X} gas discharge tube, \mbox{Ohmite 90J100E} \mbox{$100 \ \mathrm{\Omega}$} \ resistor in series with \mbox{Keithley 6482} 2-channel picoammeter
		\mbox{\circled{9} Ohmite} MOX94021006FVE \mbox{$100 \ \mathrm{M\Omega}$}\ resistors in series with Applied Kilovolts HP030RIP020~HV.
}
\end{figure*}

\subsection{Consideration of materials for new electrodes \label{subsection-electrode-material}}

We selected large-grain niobium and grade-2 titanium for testing after reviewing accelerator physics literature.
The bulk properties of these metals and other commonly used high voltage metals are cataloged in Table~\ref{bulk}.
Our goal is to use the material that sustains the highest electric field strength while minimizing leakage current and magnetic impurities that could introduce EDM systematic effects.
Stainless steel was excluded from our testing due to its relatively high ferromagnetic content but its properties are nevertheless included for reference. 
\par

Large-grain niobium electrodes with a cathode area of~3170 mm$^2$\ have been tested to fields as high as \mbox{18.7~kV/mm~\cite{Ba12}.}
Fine-grain appears to perform slightly worse, perhaps because the higher grain boundary density increases particulate adherence to the electrode surface~\cite{La06}.
The highest reported electric field for gap sizes near 1~mm that we found is 130~kV/mm using an asymmetric titanium anode and molybdenum cathode with an effective area of 7 mm$^2$~\cite{Fu05}. 
The effective area of the Ra~EDM electrode is 200 mm$^2$, approximately a factor of thirty larger. 
There is evidence that larger stressed areas are prone to lower breakdown voltages, suggesting that a miniaturized Ra~EDM electrode geometry could improve the maximum stable electric field~\cite{Phan2020}.
\par

In the presence of high electric fields, an oxide layer on an electrode surface could be a significant source of particle emission.
Niobium oxidizes at a higher rate than titanium and oxygen-free copper \cite{Sh68,Ar60,Ko61,Ge07,Og03}.
However, significant oxidation rates for these materials have only been observed at temperatures in excess of \mbox{500 $^{\circ}$C \cite{Sh68,Og03,Ge07,Su09,Gu63}.} 
The Ra~EDM experimental apparatus is pumped to ultrahigh vacuum \mbox{($< 10^{-11}$~Torr)} at room temperature.
We therefore expect that oxidation rates are negligibly low for any selection of the considered electrode materials.

We have considered a potential EDM systematic arising from magnetic impurities in the electrodes that change polarization with each electric field reversal.
A sufficiently high concentration of such impurities could perturb the magnetic field in the radium cloud region.
To address this, we measured the magnetization of copper, niobium, and titanium electrode-sized pucks in a magnetically shielded mu-metal enclosure with commercial fluxgates (Bartington Mag03IEL70).
Titanium was the most magnetic, in agreement with the magnetic properties listed in Table~\ref{bulk}. 
A custom atomic vapor cell magnetometer with a \mbox{5 mm} cube cell was also used to measure the magnetization of a pair of titanium electrodes to \mbox{$\leq 5$\ nT.}
These measurements informed our choice of large-grain niobium electrodes for the second generation EDM measurements.

\subsection{Second generation electrode surface processing \label{subsection-processing-techniques}}

We fabricated four pairs of large-grain niobium electrodes and two pairs of grade-2 titanium electrodes in two separate batches.
Surface treatment procedures for each electrode pair are cataloged in Table~\ref{electrodes} (batches 2 and 3).
\par

Our target validation field strength was 15~kV/mm or better for this phase of the Ra~EDM high voltage development.
With this in mind, we used processing procedures informed by discussions with Jefferson Lab accelerator physicists and a review of the literature. 
All but one of the second generation electrode pairs are chemically polished prior to HPR.
Recently, centrifugal barrel polishing has been shown to reduce the required conditioning time compared to chemical etching~\cite{Hernandez-Garcia2017}.
This is an encouraging prospect for conditioning Ra~EDM electrodes to significantly higher fields in a future phase of development.
\par

The four titanium electrodes (Ti$_1$, Ti$_2$, Ti$_3$, and Ti$_4$)  were mechanically polished with silicon carbide after fabrication.
Their mean surface roughness averages were measured in the range 16--23~nm using a profilometer (MicroXAM) in a clean room.
We electropolished pair $\mathrm{Ti_{13}}$\ commercially and remeasured the electrode surfaces.
We observed an increase in the surface roughness of the electropolished titanium electrodes by $\approx 50\%$\ and microprotrusions in the range 1--10~$\mu$m.
\par

We decontaminate the electrodes in clean rooms at the Facility for Rare Isotope Beams (FRIB) after polishing.
The electrodes are cleaned with detergent and rinsed with pure water in an ultrasonic bath in a staging area. 
They are rinsed in a second ultrasonic bath with UPW inside a Class~100 clean room.
The electrodes are then high pressure-rinsed with UPW at 1200 psi for twenty minutes.
After HPR, the electrodes dry in the clean room for several days before being sealed in poly tubing backfilled with dry, filtered nitrogen.
A summary of clean room and HPR parameters from several high-gradient development groups is given in Table~\ref{hpr}.

\section{\label{section-testing-the-electrodes}Electrode Discharge-Conditioning}

\subsection{High voltage test station} \label{High voltage test apparatus}

\begin{table}
\centering
\begin{threeparttable}
\caption{\label{hpr} Surface decontamination comparison.
$P = $ rinse pressure, $T =$ rinse time, CR $=$ clean room, RR $=$ rinse resistivity.}
\begin{tabular}{@{}llllll@{}}
\toprule
Lab & $P$ & $T$ & RR & CR &Ref.\\
      & (psi) & (min) & (M$\mathrm{\Omega}$ cm) & (Class) & \\
\midrule
CERN & 1500 & 30 & 18  & 100 & \cite{Be91} \\
JLab & 1200 & 20 &  $> 18$   & - & \cite{Ba12} \\
KEK  & 1100 & 5  & 80    & 100 & \cite{Fu05} \\
MSU & 1200 & 20 & 18.1 & 100 & This work
\\
\bottomrule
\end{tabular}
\end{threeparttable}
\end{table}

A schematic of the MSU high voltage test station is shown in Figure~\ref{voltage-schematic}. 
Electrode pairs are mounted to a polyether ether ketone (PEEK) holder inside a six-way cross vacuum chamber.
We estimate current flowing through the PEEK holder (resistivity \mbox{$10^{16} \ \Omega$\ cm)} is limited to less than 6~pA with an electrode voltage of 30~kV. 
The vacuum chamber is maintained at \mbox{ $10^{-7}$\ Torr} with a turbomolecular pump (Pfeiffer Hipace 80).
At this pressure the mean free path of residual gas molecules is over a meter, significantly larger than the dimensions of the chamber.
\par

The test station is frequently brought to atmospheric pressure for upgrades and electrode installations.
We perform this work in clean rooms that are validated to Class~100 or better with a NIST-calibrated particle counter (Lighthouse Handheld 3016). 
The chamber is backfilled with dry, high-purity nitrogen through a $0.01~\mu$m gas membrane particle filter (Matheson 6190 Series) while venting the chamber and after clean room operations.
During initial evacuation the pump rate is controlled at \mbox{1 Torr/s} with foreline valves to reduce the risk of disturbing vacuum chamber surfaces.\par

We use polished corona ball connections inside and outside the test chamber to minimize discharge risk beyond the electrode gap region. 
The power supply (Applied Kilovolts HP030RIP020) and feedback resistors are mounted inside a grounded high voltage cage.
The feedthroughs are enclosed by grounded ``soup can'' style shields that can be flooded with dry nitrogen to reduce humidity. \par

We use a 2-channel picoammeter (Keithley 6482) to  measure the current flowing between the electrodes.
One channel is not connected and is used to track correlated drifts between the channels.
A protection circuit between the electrode and picoammeter suppresses high-power transients that could damage the picoammeter. 
Typical discharges between the electrodes do not trigger the protection circuit.
We calibrated the picoammeter with the protection circuit to within $ 10 \ \mathrm{pA}$.

\subsection{Optical measurements of electrodes and gap sizes}

Chemical polishing removes thin layers of material from an electrode, minutely reducing its dimensions.
We developed an imaging system to measure electrode dimensions and gap sizes without making contact with the electrode.
The system uses a CMOS camera and bi-telecentric machine lens (Thorlabs MVTC23024).
\par

\begin{table}
\centering
\begin{threeparttable}
\caption{\label{daq-filter} Data acquisition and filtering settings.
Used filters are bulleted.
notch = band-rejection filter.
}
\begin{tabular}{@{}rllll@{}} \toprule
setting\hspace{2em} & Nb$_{56}$ & Nb$_{78}$  & Ti$_{13}$ & Nb$_{23}$ \\ 
\midrule
 sample rate (kHz)  & $16$ & 16 & 30 & 30 \\
  samples/point 	& $8192$ & 8192 & 8192 & 8192 \\
25--35 Hz notch   & $\hspace{0.5em}\bullet$ & $\hspace{0.5em}\bullet$ &  &  \\
 55--65 Hz notch 	& $\hspace{0.5em}\bullet$ & $\hspace{0.5em}\bullet$ &  &  \\
 109--113 Hz notch & $\hspace{0.5em}\bullet$ & $\hspace{0.5em}\bullet$ &  &  \\
 115--125 Hz notch 	& $\hspace{0.5em}\bullet$ & $\hspace{0.5em}\bullet$ &  &  \\
 7.5 kHz low-pass 	& $\hspace{0.5em}\bullet$ & $\hspace{0.5em}\bullet$ & $\hspace{0.5em}\bullet$ & $\hspace{0.5em}\bullet$ \\
\bottomrule
\end{tabular}
\end{threeparttable}
\end{table}

\begin{figure}
\centering
\includegraphics[width=0.48\textwidth]{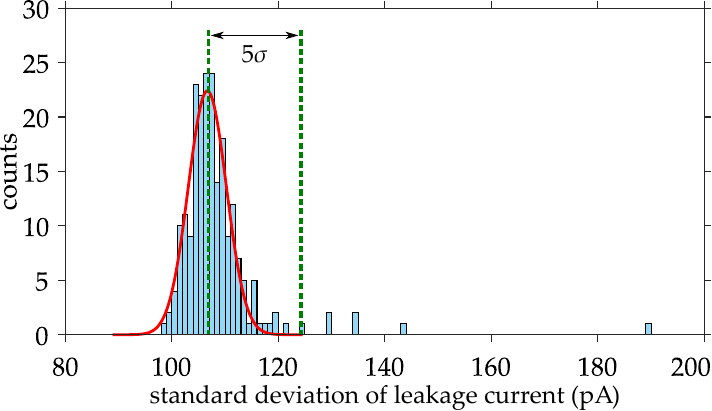}
\caption{\label{plot-gaussian-model} Nb$_{23}$\ at $+ 20$~kV/mm over one 60~second cycle during the final hour of conditioning.
From the Gaussian fit (solid red line) we determine the mean to be $\bar{x} \pm \sigma = 106.7 \pm 3.6$~pA.
There are 207 total data points.
We identified 7~events exceeding the $\bar{x} + 5\sigma = 125$~pA threshold  as discharges.
We typically observe discharge sizes in the range 100--1000~pA, but it is not uncommon to observe larger discharge sizes around 1--10~nA.}
\end{figure}
\begin{figure*}
\centering
\includegraphics[width=\drplotwidth]{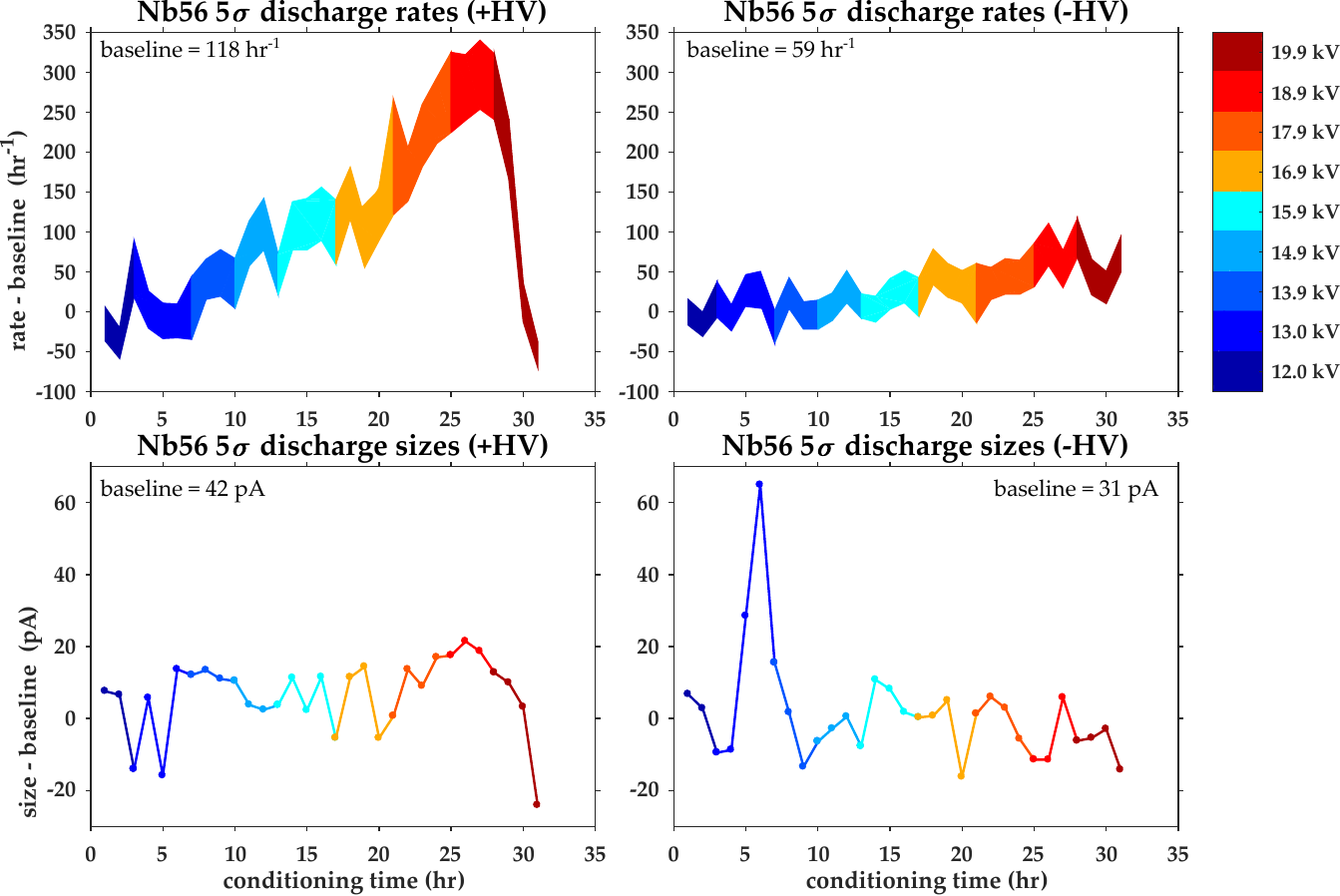}
\caption{\label{plot-discharges-nb56}
Discharge-conditioning timeline for $\mathrm{Nb}_{56}$\ at a 1 mm gap size.
}
\end{figure*}

The Ra~EDM experiment requires a gap-measuring precision of 0.1~mm or better.
To test the electrodes at different gap sizes, we adjust the gap size \textit{in situ} by translating the bottom electrode vertically with a high-precision linear drive (MDC 660002). 
We initially tested electrode performance over gap sizes ranging 0.4--2.5~mm before removing the linear drive and standardizing the gap size to \mbox{$1.0 \pm 0.1$\ mm}.
The EDM measurement features an ODT with a 50~$\mu$m waist size and requires a minimum electrode gap size of 1.0~mm to avoid heating the electrode surface.

\subsection{Data acquisition and filtering settings}

A complete description of acquisition and filtering settings used for each tested pair of electrodes is given in Table~\ref{daq-filter}. 
We record the power supply current, power supply voltage, vacuum pressure, leakage current, and rough pump foreline pressure with a 16-bit, 250 kS/s data acquisition device (NI DAQ USB-6218) connected to an office model desktop PC.
The analog signals are digitally filtered to remove 60~Hz outlet noise and mechanical vibrations from the vacuum pumps.
We initially sampled data at 16~kHz but later increased the sample rate to 30~kHz after upgrading the RAM and hard disk of the DAQ PC. 
The mean and standard deviation for each recorded data point is calculated from 8192 samples.
We removed the outlet noise filters after conditioning several pairs of electrodes because they introduced artificial shapes in the signal waveform.
Comparing the leakage current data of electrode pairs with different filtering settings, we found that the digital filters did not significantly affect the distribution of the dataset discussed in Section~\ref{subsection-identify-discharges}.
We are sensitive to absolute currents as small as \mbox{$\sigma \approx 25$\ pA} with the acquisition settings described in Table~\ref{daq-filter}. 

\subsection{Identifying electrode discharges}\label{subsection-identify-discharges}

We observe discharges on a timescale of approximately 2~ms. 
Each sample mean and standard deviation is calculated over an integration period of \mbox{$\approx 270$\ ms}.
The sample mean is well suited for characterizing the relatively slow steady-state leakage current.
On the other hand, we find that the sample standard deviation is effective for counting discrete discharges and estimating discharge size.
\par
The standard deviation of the leakage current monitor signal is derived from 8192 samples over a time period.
For a 30~kHz sample rate, this corresponds to a total time period of 273~ms. 
About 220 leakage current standard deviations are collected over the 60 second time period that corresponds to one high voltage magnitude and polarity setting. 
\par
\begin{figure*}
\centering
\includegraphics[width=\drplotwidth]{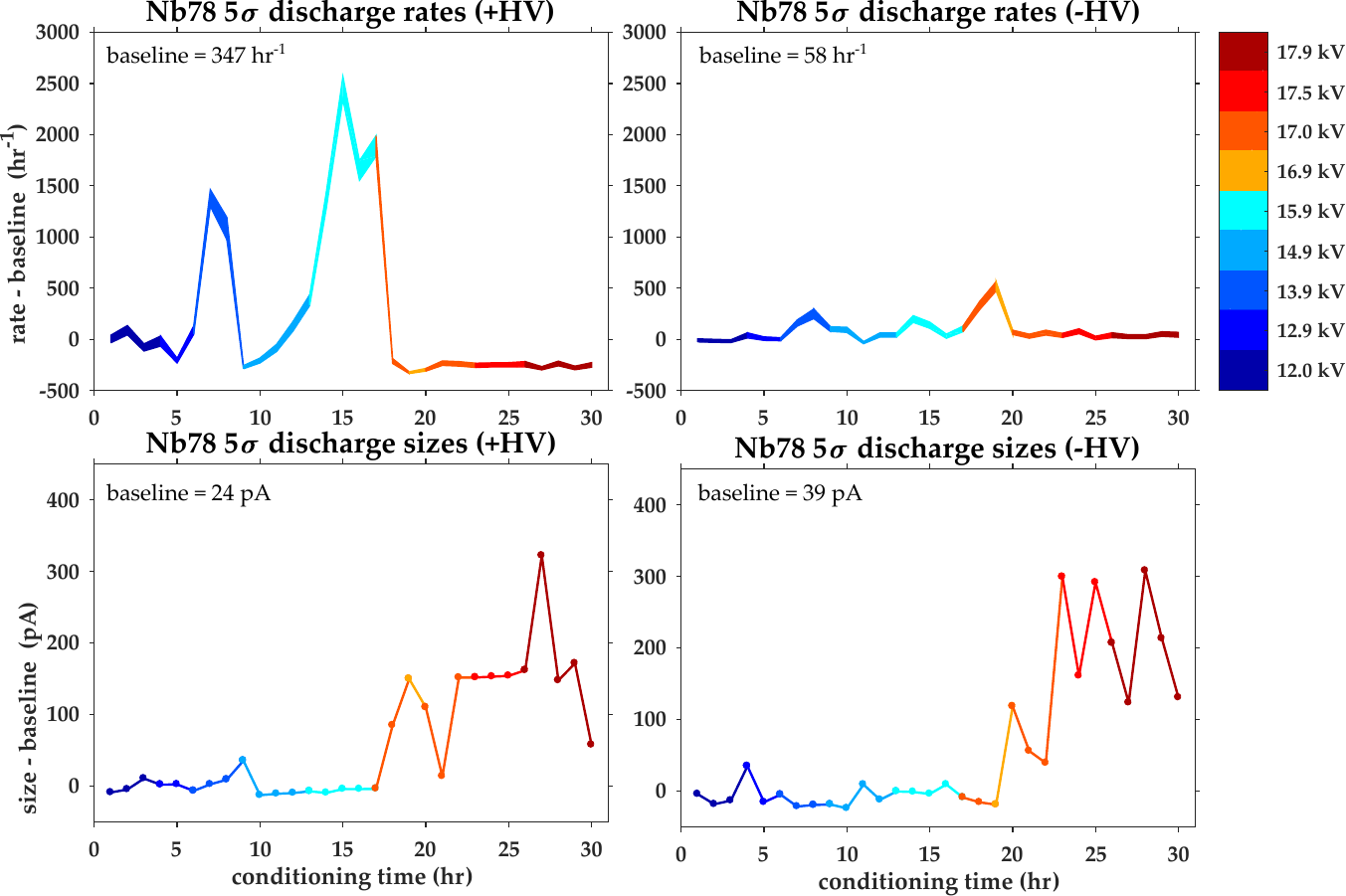}
\caption{\label{plot-discharges-nb78}\footnotesize{
Discharge-conditioning timeline for $\mathrm{Nb}_{78}$\ with a 1 mm gap size. 
}}
\end{figure*}
The distribution of these 220 leakage current standard deviations is reasonably modeled as a Gaussian distribution.
An example of one set of data over a 60~s time period is given in Figure~\ref{plot-gaussian-model}. 
Using the mean and standard deviation of this distribution, we set a threshold that is five standard deviations above the mean. 
We define a ``discharge" as any leakage current standard deviation above this threshold. 
\par
\begin{figure*}
\centering
\includegraphics[width=\drplotwidth]{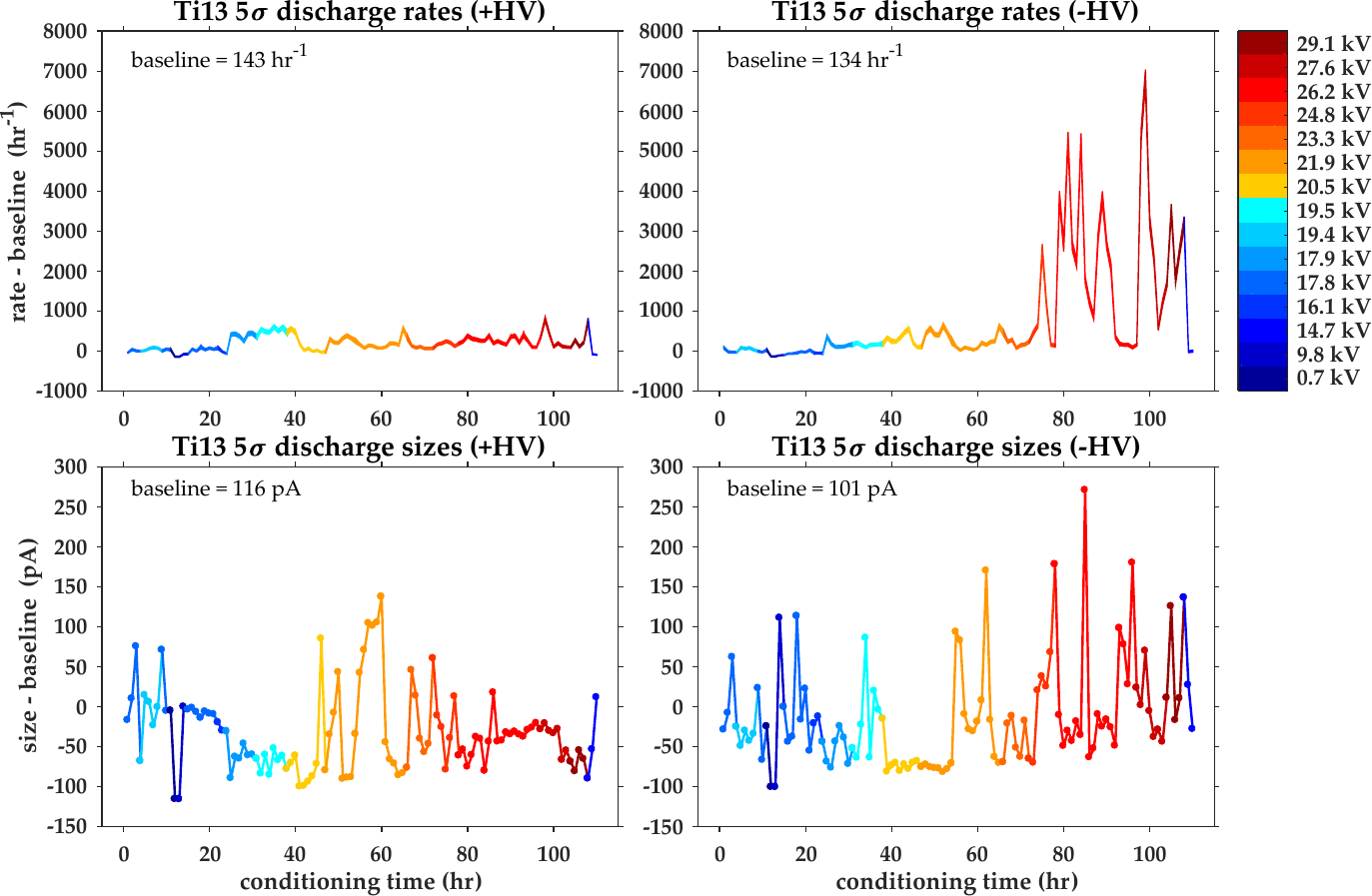}
\caption{\label{plot-discharges-ti13}\footnotesize{
Discharge-conditioning timeline for $\mathrm{Ti}_{13}$\ at a 0.9 mm gap size. 
}}
\end{figure*}

Discharge rates and typical discharge sizes are presented in \mbox{Figures \ref{plot-discharges-nb56}, \ref{plot-discharges-nb78}, \ref{plot-discharges-ti13}, and \ref{plot-discharges-nb23}.} 
A constant ``baseline'' is subtracted from each plot.
We define the baseline as the discharge rate and discharge size at the high voltage magnitude at the start of the conditioning process.
\par

To estimate discharge magnitudes, we report the median value of the standard deviations in each 60~s time period. 
We expect to see high rates of discharges during discharge-conditioning. 
Small discharges occurring at a stable rate are beneficial and do not damage the electrode surfaces.
\par
Our discharge counting method does not exclude discharges that could occur at another part of the test station, for example the high voltage feedthroughs.
Therefore, we expect our reported discharge rates are conservative overestimates of the true electrode discharge rate.  
\par

Additionally, we calculate the steady-state leakage current $\bar{I}$\ using the Gaussian mean of each $273$~ms time period.
$\bar{I}$\ is insensitive to discharges, which typically last $\approx 1$\ ms.
For example, in the third hour of conditioning Nb$_{56}$\ at 19.9~kV (see Figure~\ref{plot-discharges-nb56}), we count a polarity-combined  54 events above the $5\sigma$\ threshold with the standard deviations but only 2~events with the mean data during the same period.
\par

Our analysis code models the leakage current and calculates the discussed performance metrics.
The accuracy of the analysis was independently verified. 
Code and data availability may be found in \ref{appendix-code}.

\subsection{Discharge-conditioning procedure}

Our goal is maximize the electric field strength while minimizing the discharge rate and discharge size.
We condition the electrodes at DC voltage, alternating the voltage polarity every 60~s.
The voltage is applied to the top electrode.
The periodic voltage waveform is chosen to simulate the EDM measurement and is more challenging to stabilize than holding off a static unipolar field.
We usually observe the highest rates of discharges during the second and third hours of conditioning. 
\par

In the final conditioning phase we validate the electrodes at some fraction of the maximum voltage to reduce the discharge rate.
The validation voltage is typically $80$--$95\%$\ of the maximum tested voltage~\cite{La95,La06}. 
\par

In Sections~\ref{nb56-analysis}, \ref{nb78-analysis}, \ref{ti13-analysis}, and \ref{subsxn-nb23-rates}, we will discuss the discharge-conditioning results of each electrode pair.
In Section~\ref{subsection-overall-comparison}, we will compare the overall electrode performance.

\subsection{$\mathrm{Nb}_{56}$\ conditioning results \label{nb56-analysis}}

The average discharge rate over the course of conditioning the niobium electrode pair $\mathrm{Nb}_{56}$ is shown in the upper panels of Figure~\ref{plot-discharges-nb56}.
At each voltage, the discharge rates, expressed in discharges per hour (dph), tend to decrease as we condition. 
There is a step-like increase in discharge rates when the voltage is increased.
$\mathrm{Nb}_{56}$\ was validated at \mbox{20 kV / 1 mm} with an average discharge rate of \mbox{$98 \pm 19$\ dph} after approximately thirty hours of conditioning.
\par

At negative polarity, the discharge rate increases more slowly with each voltage step.
However, the overall curve does not flatten at a minimum count rate as it does at positive polarity. 
This suggests that additional conditioning could further suppress discharges at negative polarity.
It is also possible that the test station design facilitates a higher discharge rate at negative polarity. 
We will explore this in the near future by conducting conditioning tests while the electrodes are removed from the test station.
\par

Nb$_{56}$\ discharge sizes are shown in the lower panels of Figure~\ref{plot-discharges-nb56}.
As we will see with all the discharge plots, the discharge size behavior does not scale with the discharge rate.
The largest median discharge size over the course of conditioning is 60 pA, which is relatively small compared to the typical discharge sizes of the other electrode pairs. 
In the last hour of conditioning the discharge sizes are 20~pA smaller than the starting discharge sizes. 
\par

\begin{figure*}
\centering
\includegraphics[width=\drplotwidth]{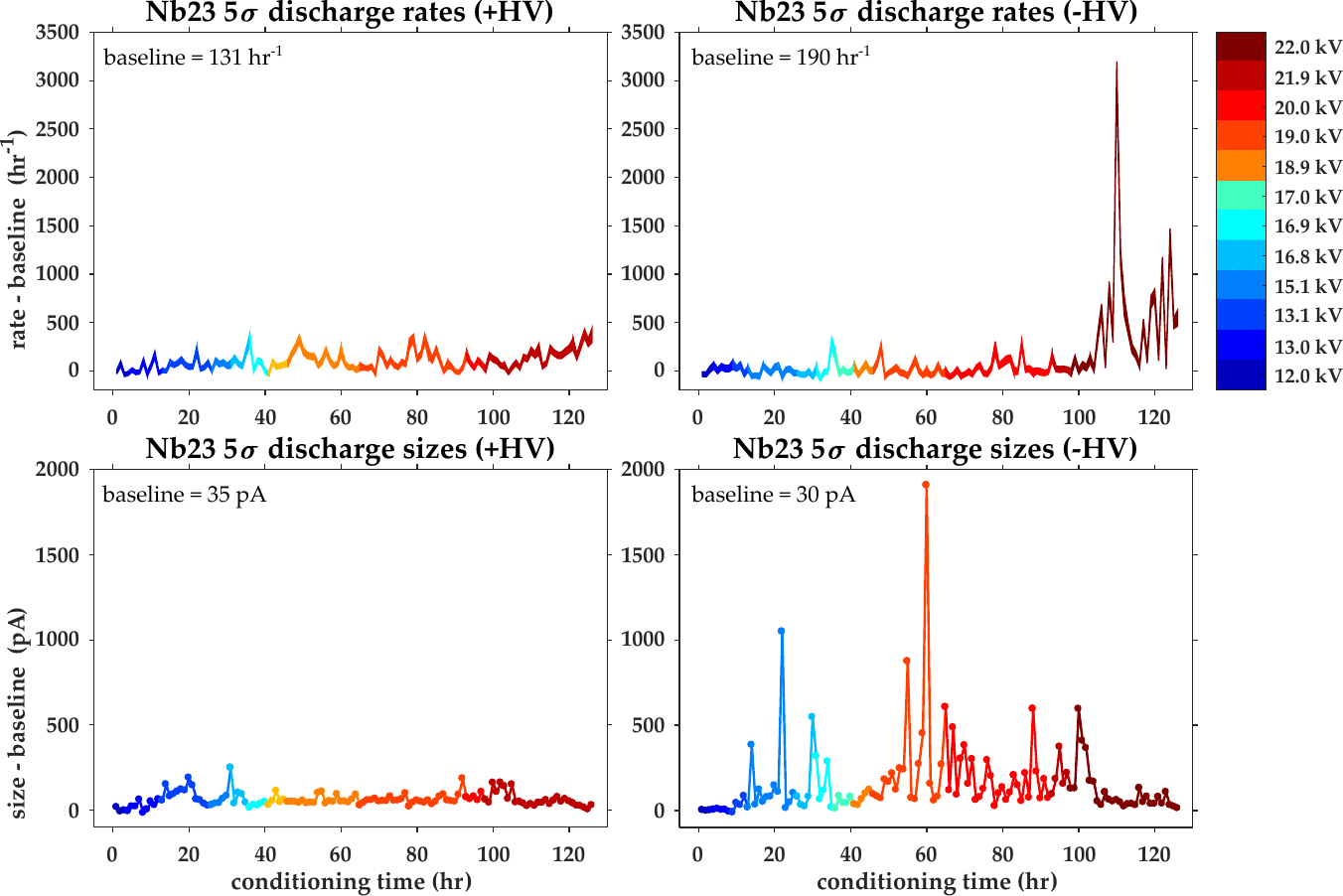}
\caption{\label{plot-discharges-nb23}\footnotesize{
Discharge-conditioning timeline for $\mathrm{Nb}_{23}$\ at a 1 mm gap size. 
}}
\end{figure*}

As mentioned in Section~\ref{early-anl}, the legacy copper electrodes were conditioned to 10~kV/mm but could only be operated at 6.5~kV/mm after installing them in the Ra~EDM apparatus. 
For the second generation electrodes, we made two major improvements to our technique to prevent a similar reduction in field strength.
First, our electrodes are now preserved in Class~100 or better clean room environments during conditioning and transport as described in \mbox{Sections \ref{subsection-processing-techniques} and \ref{High voltage test apparatus}}.
Second, we used the new, rigorous discharge-conditioning procedure described Section~\ref{subsection-identify-discharges} for Nb$_{56}$\ and the electrodes discussed in the subsequent sections.
\par

Nb$_{56}$\ was installed in the Ra~EDM apparatus using the clean room methods described in Section \ref{section:strategy-results}.
They were revalidated at 20 kV/mm after installation.
This electrode pair will be used for upcoming second generation EDM measurements.

\subsection{$\mathrm{Nb}_{78}$\ conditioning results\label{nb78-analysis}}

Discharge rates and sizes for the second pair of niobium electrodes Nb$_{78}$\ are given in Figure~\ref{plot-discharges-nb78}.
We started conditioning Nb$_{78}$\ at 12~kV/ 1 mm, the same electric field as Nb$_{56}$.
The initial discharge rates are occasionally in excess of 1000 dph, or about once every three seconds for several hours with discharge sizes of 50~pA.
The high discharge rate coupled with low discharge size is an indication that we are operating at an optimized voltage for discharge-conditioning.
During the last 10 hours of conditioning the discharge rates decrease to less than the initial rates.
The final conditioning shift was performed at 17.8 kV/mm.
\par

These electrodes were packaged according to our procedure described in Section \ref{section:strategy-results} and shipped to the University of Science and Technology of China, where they are being used in an ytterbium EDM measurement.

\subsection{$\mathrm{Ti}_{13}$\ conditioning results\label{ti13-analysis}}

We changed our data acquisition and digital filter settings for Ti$_{13}$ and the pair that we will discuss in Section~\ref{subsxn-nb23-rates} (see Table~\ref{daq-filter}).
To reach  electric fields higher than \mbox{$20$\ kV/mm,} we conditioned the titanium electrodes for \mbox{$\approx 110$}\ hours, four times longer than the previous pairs. 
\par

Discharge rates and sizes for the titanium electrodes are shown in Figure~\ref{plot-discharges-ti13}. 
We started conditioning the electrodes at 14.9 kV/ 0.9 mm = 16.5 kV/mm.
The initial discharge sizes are approximately 100~pA, significantly higher than Nb$_{56}$\ and Nb$_{78}$.
The discharge rates did not consistently decrease over the course of several shifts at 19.4~kV.
At hour~12, we reduced the voltage to 0.7~kV for one shift to verify that the discharge rates decrease before resuming testing at higher voltages. 
\par

The discharge rate increases from 290~dph to 5550~dph when stepping the voltage from $-26.2$~kV to $-27.6$~kV. 
This step-like `switching on' of leakage emission sites is consistent with our expectations, given the physical picture of conditioning we describe in Section~\ref{section:strategy-results}.
In principle, the emission sites, which may be thought of as microprotrusions, are ablated after spending sufficient time discharge-conditioning the electrodes.
The factors influencing the required amount of time include the smoothness of the high-gradient surfaces, the gap size, and the applied voltage.
We were unable to significantly reduce the discharge rates at \mbox{27.6 kV / 0.9 mm =  30.7 kV/mm} despite more than twenty hours of conditioning. 
\par

During the final shift, we reduced the voltage to \mbox{14.7~kV~/~0.9 mm =~16.3~kV/mm} and again observed the discharge rates returning to the baseline.
Ti$_{13}$\ can likely be conditioned to perform stably at $\approx 24 $~kV, or 85\% of the maximum applied voltage with additional conditioning. 
However, the concentration of magnetic impurities in our titanium electrodes (shown in Table \ref{bulk}) is likely too high to be used for an EDM measurement.

\subsection{$\mathrm{Nb}_{23}$\ conditioning results\label{subsxn-nb23-rates}}

We first tested Nb$_{23}$\ at a 0.4~mm gap with fields as high as \mbox{$+52.5$~kV/mm} and \mbox{$-51.5$~kV/mm} using the traditional hold-off or ``current-conditioning'' method~\cite{La95}.
Then we discharge-conditioned the electrodes with the periodic waveform described in Section~\ref{subsection-identify-discharges} to 27.5~kV/mm.
During the subsequent 30~kV/mm conditioning shift, an approximately 100~nA discharge  triggered a current avalanche that rapidly increased the leakage current and damaged the electrodes.
We were unable to recover meaningful performance with discharge-conditioning and repolished the surface according to Table~\ref{electrodes}. \par

Repolished Nb$_{23}$\ discharges rates are shown in Figure~\ref{plot-discharges-nb23}.
The rates stay near the baseline, about 200~dph for both polarities up to 20~kV.
When we increased the voltage from \mbox{20 to 22 kV,} the discharge rates become as high as 3000~dph (about once every second).
The discharge sizes were low, less than 500~pA, so we continued conditioning at this voltage.
Despite conditioning the electrodes at 22~kV/mm for more than twenty hours, the discharge rate remained high. 
We expect that reducing the voltage by \mbox{$\approx 1$\ kV} will restore the baseline discharge rate.
\par

As noted previously, we were conditioning Nb$_{23}$\ at 30~kV/mm before a destructive discharge inhibited performance.
We recovered 80\% of the original electric field performance by repolishing and reconditioning Nb$_{23}$.

\begin{table}
\centering
\begin{threeparttable}
\caption{Electrode conditioning summary.
\mbox{$E_i  =$\ initial} field strength.
\mbox{$E_{\mathrm{max}} =$\ max} field strength.
\mbox{$E_{f} =$\ final} validated field strength.
\mbox{$R_i \ (R_f)  =$\ initial} (final) discharge rate.
\mbox{$\bar{I} =$\ steady}-state current at $E_f$. 
} 
\label{relative-electrode-performance}
\begin{tabular}{@{}lccccc@{}} 
\toprule
\multirow{2}{*}{pair} & $E_{i}$ & $E_{\mathrm{max}}$    & $E_{f}$ & \multirow{2}{*}{$\dfrac{R_{f}}{ R_{i}}\ $} &  $\bar{I}$ \\
				  & (kV/mm)  & (kV/mm) & (kV/mm) &  &    (pA) \\
\midrule
$\mathrm{Nb}_{56}$ & 11.9 & 19.8 & 19.8 & 1.6 &  $< 10$ \\ 
$\mathrm{Nb}_{78}$ & 12.0 & 17.9 & 17.9 & 0.9 &  $< 10$\\ 
$\mathrm{Ti}_{13}  $ & 19.8 & 32.3 & 29.1 & 2.2 &  $< 30$ \\ 
$\mathrm{Nb}_{23}$ & 12.0 & 22.0 & 21.9 & 1.3 &  $< 25$ \\
\bottomrule
\end{tabular}
\end{threeparttable}
\end{table}

\subsection{Comparison of overall electrode performance\label{subsection-overall-comparison}}

Table~\ref{relative-electrode-performance} compares the electric fields tested and discharge rates observed for all of the conditioned electrode pairs averaged over both polarities.
$E_i$\ is the electric field strength at the start of the conditioning, while $E_\mathrm{max}$\ is the electric field strength at the end of the conditioning. 
$E_i$\ was chosen, based on the best performance reported in the literature, to be sufficiently low to minimize the possibility of a catastrophic discharge while sufficiently high to minimize the conditioning time. 
$E_f$\ is the final validated field strength, and can be some percent lower than $E_\mathrm{max}$. 
This is a customary margin which helps ensure stable performance~\cite{La95,La06}. 
\par

All the electrodes were discharge-conditioned to higher than our original goal of 15~kV/mm. 
The steady-state currents $\bar{I}$\ at $E_f$\ are under the EDM experimental threshold of 100~pA.
The ratio of the discharge rate at the final field strength to the initial field strength $R_f/R_i$\ are within a factor of 2.2.
In the case of Nb$_{78}$, the final average discharge rate is lower than the initial discharge rate.
We conditioned Ti$_{13}$ most aggressively and it has both the highest final field strength and largest relative increase in discharge rate.
 \par

Of particular note is the polarity dependence of the electrode discharge rates. 
In all cases except for Nb$_{23}$, the negative polarity discharge rates are significantly higher than the discharge rates at positive polarity.
Polarity-dependent discharge rates could be a feature of permanently grounding the bottom electrode and only charging the top electrode, as illustrated by Figure~\ref{voltage-schematic}.
In the future, we plan to design a more symmetric test station that will alternate the role of grounded and charged electrode to further investigate this effect.
\par

We calculated the steady-state leakage current for all conditioning runs by subtracting the Gaussian means of zero voltage periods from those of high voltage periods.
This removes leakage offsets and drifts due to the picoammeter, protection circuit, and power supply.
The steady-state leakage current versus voltage trend is modestly linear with an ohmic resistance of  \mbox{$40 \ \mathrm{kV} / 10 \ \mathrm{pA} \approx 10^{16} \ \Omega$.}
We observe large leakage currents \mbox{$ > 100$\ pA,} correlated with high discharge rates, for Ti$_{13}$\ and Nb$_{23}$\ beyond 22~kV. 
In the most extreme case, we measured \mbox{$\bar{I} \approx -670$~pA} during conditioning Ti$_{13}$\ at $-27.6$~kV.
\par

The steady-state leakage current must be less than \mbox{100 pA} to avoid systematics that could mimic an EDM signal at our current statistical sensitivity.
This criterion is similar to metrics used in other electrode development groups~\cite{Ba12,Fu05}.
Our steady-state leakage current sensitivity is limited to \mbox{$\approx25$\ pA.}
At the Nb$_{56}$\ final validated field strength of 20~kV/mm, the measured steady-state leakage current is below this upper limit.

\section{Conclusions and outlook\label{section-conclusions}}

The Ra~EDM experiment searches for the atomic electric dipole moment of $^{225}$Ra.
During the measurement, the atom spins precess between a pair of identical plane-parallel electrodes that generate a uniform and stable DC electric field that reverses direction every measurement cycle. 
We used a pair of oxygen-free copper electrodes that operated at \mbox{$\pm 6.7$\ kV/mm} and measured an EDM upper limit of \mbox{$1.4\times 10^{-23}\ e$ cm} in the first generation of measurements.
For the second generation measurements, we will use a new pair of large-grain niobium electrodes whose systematic effects have been evaluated to the \mbox{$10^{-26}\ e$ cm} level.
\par

Two pairs of grade-2 titanium and four pairs of large-grain niobium electrodes were fabricated and polished according to surface preparation techniques that we adapted from accelerator physics literature.
We constructed a high voltage test station to condition high voltage electrodes at gap sizes of \mbox{0.4--2.5 mm} with a \mbox{$30 \ \mathrm{kV}$} bipolar power supply at MSU.
Procedures were developed to decontaminate electrodes and preserve them in Class~100 environments.
\par

We discharge-conditioned three pairs of niobium electrodes and one pair of titanium electrodes, alternating the polarity of the applied DC field every \mbox{60 s} to mimic the EDM measurement.
Electric fields were tested as high as \mbox{$+52.5$\ kV/mm} and \mbox{$-51.5$\ kV/mm.}
All the electrodes exhibited less than 100~pA steady-state leakage current when operated under \mbox{22 kV.}
We validated a pair of large-grain niobium electrodes at \mbox{20 kV/mm} with an average discharge rate of \mbox{$98 \pm 19$\ discharges per hour} and a steady-state leakage less than 25~pA ($1\sigma$).
\par

The large-grain niobium electrodes ($\mathrm{Nb}_{56}$) were transported to ANL and installed in the \mbox{Ra EDM} apparatus all while preserving the electrodes in Class~100 environments.
After installation, the performance of $\mathrm{Nb}_{56}$\ was revalidated at 20~kV/mm.
The improved electric field strength is expected to contribute a factor of 3.1 increase in our EDM statistical sensitivity. 
\par
In the next phase of the Ra~EDM high voltage development, we will design a more symmetric high voltage test chamber using a unipolar power supply that alternates the field direction by switching connections between the electrodes.
Our plan is to discharge-condition electrodes to operate reliably at \mbox{$\pm 50$~kV/mm} across a \mbox{1 mm} gap.

\section*{Acknowledgments}

The authors would like to thank: Matthew Poelker and the Electron Gun group at Jefferson Lab for fabricating our electrodes and advising on polishing techniques; TU Munich (TUM) for sharing their mu-metal prototype enclosure for our electrode magnetization measurements; Zheng-Tian Lu and his EDM group at the University of Science and Technology of China for sharing their electrode magnetization measurements; Laura Popielarski and Daniel Victory (FRIB) for helping us high-pressure rinse our electrodes; and Samuel Nash (NSCL) for advising us on clean room design and validation. \par

We acknowledge support from: Michigan State University; US DOE Office of Science, Office of Physics under DE-AC02-06CH11357; DOE Oak Ridge Institute for Science and Education DE-SC0014664; DOE National Nuclear Security Administration through NSSC DE-NA0003180; and US DOE Office of Science, Office of Nuclear Physics under contract DE-SC0019455.

\appendix

\section{\label{mjn-calc}Magnetic Johnson noise}

One source of magnetic field instability that could potentially limit the sensitivity of this experiment is the Johnson-Nyquist noise \cite{johnson,nyquist} 
from conducting materials near the detection region.
Thermal agitation (i.e. energy fluctuations) of the charge carriers inside conductors give rise to this electronic noise with a nearly frequency-independent spectral power density of:
\begin{equation}
	\frac{d P_n}{d \nu} = 4 k_{\mathrm{B}} T
\end{equation}
where $k_\mathrm{B}$\ is the Boltzmann constant and $T$ is the temperature in Kelvin.
A derivation of this equation as well as a discussion of its frequency dependence is given in \cite{abbott}. 
By noting that the power dissipated by a conductor is given by $P = I^2 R$,  we can rewrite the noise spectrum in terms of the RMS current noise as:

\begin{equation}
\sqrt{I_n^2} 
\ \ \   
= 
\ \ \    
\sqrt{\frac{dI_n^2}{d\nu} (\Delta \nu)} 
\ \ \    
= 
\ \ \  
\sqrt{\frac{4 k_{\mathrm{B}} T (\Delta \nu)}{R}} \label{eqn:irms}
\end{equation}

\noindent where $R$ is the resistance and $\Delta \nu$ is the bandwidth.
This current noise generates a magnetic field noise spectrum that, in general, depends on the geometry of and distance from the conductor and the frequency.

In general, it is quite onerous to follow the frequency-dependent prescription of Varpula \& Poutanen~\cite{vp1984} for arbitrary geometries.  
However, as pointed out by Lamoreux \cite{lamo}, 
calculating the noise density at zero frequency always provides a conservative upper limit for the noise density at all frequencies.
In this case, called the quasistatic case, we ignore the effect of eddy currents and are able to directly apply the Biot-Savart Law to calculate 
the magnetic field from a steady state current distribution:

\begin{equation}
	d\vec{B}(\vec{r}) = \frac{\mu_0}{4\pi} \left [ \frac{I d\vec{\ell} \times \left ( \vec{r} - \vec{u} \right )  }{|\vec{r}-\vec{u}|^3} \right ] 
\end{equation}

\noindent where $\vec{B}(\vec{r})$ is the magnetic field at the location $\vec{r}$,
$I$ is the current, 
and $d\vec{\ell}$ is the line element in the direction of the current at the location $\vec{u}$.
This integral over $d\vec{\ell}$  is assumed to be zero for randomly fluctuating noise currents.

On the other hand, the RMS magnetic field is not expected to be zero and, for example, the $y-$component can be written as:

\begin{equation}
\label{eqn:db2}
\begin{split}
	dB^2_y &= \frac{\mu_0^2}{16\pi^2} \left [ \frac{\left \{ I_x d \ell_x \left ( r_z - u_z \right )  - I_z d\ell_z \left ( r_x - u_x \right ) \right \}^2  }{|\vec{r}-\vec{u}|^6} \right ]  
  \\
	  &= \frac{\mu_0^2}{16\pi^2} \left [ \frac{ I_x^2 (d \ell_x)^2 \left ( r_z - u_z \right )^2  + I_z^2 (d\ell_z)^2 \left ( r_x - u_x \right )^2}{|\vec{r}-\vec{u}|^6} \right ]
\end{split}
\end{equation}

\noindent where $I_q$ is the current in the $q$ direction and the subscripts $q=x,y,z$ label the component of the vectors such that $\hat{x} \times \hat{y} = \hat{z}$. 

The randomly fluctuating noise currents in two different directions are assumed to be completely uncorrelated.
Therefore, the cross term (i.e. $I_x I_z$) is assumed to integrate to zero and only the quadratic terms (i.e. $I_x^2$, $I_z^2$) survive.
The field noise density can be written in terms of the current noise density, which, in the $q$ direction, is given by:

\begin{equation}
\frac{dI_{n,q}^2}{d\nu} = \frac{4 k_{\mathrm{B}} T}{R_q} = \frac{4 k_{\mathrm{B}} T}{\rho} \left ( \frac{d A_q}{d \ell_q} \right ) 
\end{equation}

\noindent where $R_q$ is the resistance in the $q$ direction, $d \ell_q$ is the length in the $q$ direction, and $d A_q$ is the cross sectional area normal to the $q$ direction. 
For example, for a randomly fluctuating current in the $x$-direction, we have:
\begin{equation}
\frac{dI_{n,x}^2}{d\nu} = \frac{4 k_{\mathrm{B}} T}{R_x} = \frac{4 k_{\mathrm{B}} T}{\rho} \left ( \frac{d A_x}{d \ell_x} \right ) = \frac{4 k_{\mathrm{B}} T}{\rho} \left ( \frac{d \ell_z d \ell_y}{d \ell_x} \right )\end{equation}
Plugging this into Eqn.~(\ref{eqn:db2}) and dropping the cross terms (as argued before), we find that the $q$ component of the field noise density is given by:
\begin{equation}
	\frac{d B^2_{n,q}}{d \nu} = \left ( \frac{\mu_0^2 k_{\mathrm{B}} T}{4\pi^2\rho } \right ) \int \left | \frac{\left ( \vec{r}-\vec{u} \right )\times \hat{q} }{|\vec{r}-\vec{u}|^3} \right |^2 d^3u 
\label{eqn:Bnoiseint}
\end{equation}
\noindent where $d^3 u = (d \ell_x)(d \ell_y)(d \ell_z)$ and the scale factor is
\begin{equation}
 \frac{\mu_0^2 k_{\mathrm{B}} T}{4\pi^2\rho } = 
 \left [ \frac{0.989\ \mathrm{pT}}{\sqrt{\mathrm{Hz}}} \right ]^2 
\left [  \frac{T}{273\ \mathrm{K}} \right ] \left [ \frac{\rho_\mathrm{Cu}(273\ \mathrm{K})}{\rho(T)}  \right ] \cdot \mathrm{cm} 
\end{equation}
\noindent where $\rho_\mathrm{Cu}(273\ \mathrm{K}) = 1.542\!\times\!10^{-8}\ \Omega \cdot \mathrm{m}$.

For an infinite conducting plane of thickness $d$, Varpula \& Poutanen \cite{vp1984} have found an analytic form for the noise volume integral:
\begin{equation}
\int \left | \frac{\left ( \vec{r}-\vec{u} \right )\times \hat{y} }{|\vec{r}-\vec{u}|^3} \right |^2 d^3u = \frac{\pi}{2y} \left [ \frac{d}{d+y} \right ]
\end{equation}
\noindent where $y$ is the distance from the surface of the conducting plane.
While a numerical integration of the noise volume integral for the noise in the $y$-direction converges to the analytic formula above, the same numerical integration suggests that, contrary to the conclusion of Varpula \& Poutanen, the three components of the magnetic field noise due to an infinite plane are related by:
\begin{equation}
\left ( \frac{dB^2_{n,x}}{d\nu} \right ) = \left ( \frac{dB^2_{n,z}}{d\nu} \right ) = \frac{3}{2} \left ( \frac{dB^2_{n,y}}{d\nu} \right ) 
\end{equation}

Finally, we model the two Ra~EDM HV electrodes as cylinders with radius of 1.2 cm and height of 1.6 cm.
Assuming a gap between electrodes of 1 mm, noise volume integrals at a location directly in between the electrodes are:

\begin{equation}
	\begin{split}
		\int \left | \frac{\left ( \vec{r}-\vec{u} \right )\times \hat{x} }{|\vec{r}-\vec{u}|^3} \right |^2 d^3u  &= \int \left | \frac{\left ( \vec{r}-\vec{u} \right )\times \hat{z} }{|\vec{r}-\vec{u}|^3} \right |^2 d^3u \\
		  &=  92.5 \ \mathrm{cm}^{-1} \\
		\int \left | \frac{\left ( \vec{r}-\vec{u} \right )\times \hat{y} }{|\vec{r}-\vec{u}|^3} \right |^2 d^3u &= 56.8 \ \mathrm{cm}^{-1}
	\end{split}
\end{equation}

For Niobium electrodes ($\rho_\mathrm{Nb}/\rho_\mathrm{Cu} = 9.85$) that are held at room temperature ($T=298\ \mathrm{K}$), we calculate a magnetic field noise density of:
\begin{equation}
	\begin{split}
\sqrt{\frac{dB^2_{n,x}}{d\nu}} = \sqrt{\frac{dB^2_{n,z}}{d\nu}} &= 3.17 \mathrm{\frac{pT}{\sqrt{Hz}}}  \\
\sqrt{\frac{dB^2_{n,y}}{d\nu}} &= 2.48 \mathrm{\frac{pT}{\sqrt{Hz}}} 
	\end{split}
\end{equation}

\section{Code and data availability \label{appendix-code}}

The code used to analyze the high voltage data and generate the current discharge plots is available for use at \href{https://zenodo.org/badge/latestdoi/294766922}{https://zenodo.org/badge/latestdoi/294766922}.
The data used for the high voltage analysis may be made available for reasonable requests sent to \Verb|singhj@frib.msu.edu|.

\bibliography{hv-conditioning-msu}{}
\end{document}